\documentclass[twocolumn,,amsmath,amssymb]{revtex4}
\usepackage{graphicx}
\usepackage{dcolumn}
\usepackage{bm}
\newcommand{\etal}{\emph{et al.}}
\newcommand{\be}{\begin{equation}}
\newcommand{\ee}{\end{equation}}
\newcommand{\bfig}{\begin{figure}}
\newcommand{\efig}{\end{figure}}

\begin{document}      
\title{A pressure-induced topological phase with large Berry curvature in Pb$_{1-x}$Sn$_x$Te 
} 

\author{Tian Liang$^1$, Satya Kushwaha$^2$, Jinwoong Kim$^3$, Quinn Gibson$^2$, Jingjing Lin$^1$, Nicholas Kioussis$^3$, R. J. Cava$^2$ and N. P. Ong$^1$}
\affiliation{
{Departments of Physics$^1$ and Chemistry$^2$, Princeton University, Princeton, New Jersey 08544}\\
$^3$Department of Physics and Astronomy, California State University Northridge, Northridge, California 91330
} 

\date{\today}      
\begin{abstract}
The picture of how a gap closes in a semiconductor has been radically transformed by topological concepts.
Instead of the gap closing and immediately re-opening, topological arguments predict that, in the absence of inversion symmetry, a metallic phase protected by Weyl nodes persists over a finite interval of the tuning parameter (e.g. pressure $P$) . The gap re-appears when the Weyl nodes mutually annihilate. We report evidence that Pb$_{1-x}$Sn$_x$Te exhibits this topological metallic phase. Using pressure to tune the gap, we have tracked the nucleation of a Fermi surface droplet that rapidly grows in volume with $P$. In the metallic state we observe a large Berry curvature which dominates the Hall effect. Moreover, a giant negative magnetoresistance is observed in the insulating side of phase boundaries, in accord with \emph{ab initio} calculations. The results confirm the existence of a topological metallic phase over a finite pressure interval.
\end{abstract}
 
\maketitle      

\noindent
{\bf Introduction}\\
\noindent
Topological concepts have greatly clarified the role of symmetry in protecting electronic states in a host of materials. In bulk semiconductors, topological insights have revised the picture of how the energy gap closes (say, under pressure $P$). In the old picture, the gap $\Delta$ closes at an ``accidental'' value of $P$ before reopening at higher $P$. The new view~\cite{Murakami1,MurakamiKuga,Okugawa,YangNagaosa} predicts instead that, when inversion symmetry is broken, a gapless metallic state featuring pairs of Weyl nodes persists over a finite interval ($P_1\to P_2$). They act as sources and sinks of Berry curvature (an effective magnetic field in $\bf k$ space). The metallic phase is protected because the nodes come in pairs with opposite chiralities ($\chi = \pm 1$). Hence they cannot be removed except by mutual annihilation (which eventually occurs at the higher pressure $P_2$). To date, these predictions have not been tested.

Here we show that Pb$_{1-x}$Sn$_x$Te exhibits a pressure-induced metallic phase described by the Weyl scenario. The Pb-based rocksalts~\cite{Wallis,Dimmock,Akimov} have been identified as topological crystalline insulators with surface states protected by mirror symmetry~\cite{Fu,Hsieh,Story,Takahashi,Hasan}. Here we focus on their Dirac-like bulk states~\cite{Madhavan,TianLiang} which occur at the $L$ points of the Brillouin Zone (BZ) surface. Pb$_{1-x}$Sn$_x$Te exhibits an insulator-to-metal (IM) transition at $P\sim$ 10 kbar~\cite{Akimov}. However, the IM transition is little explored. We report that the metallic state appears by nucleating 12 small Fermi Surface (FS) nodes. The breaking of time reversal symmetry (TRS) in applied $\bf B$ leads to a large Berry curvature $\mathbf{\Omega}$. Finally, we also observe an anomalously large, negative magnetoresistance (MR) which is anticipated in the Weyl scenario.

\vspace{5mm}
\noindent
{\bf Results \\
\noindent
Phase diagram under pressure}\\
\noindent
Crystals of Pb$_{1-x}$Sn$_x$Te were grown by the vertical Bridgman technique (see Materials and Methods). Transport measurements at temperatures $T$ down to 2 K were carried out in a Be-Cu pressure cell (with a maximum pressure $P_{max}\sim$ 28 kbar) on 4 samples with Sn contents $x$ = 0.5 (Sample A1, A2), 0.32 (Q1) and 0.25 (E1) (see Table~\cite{SI}). In A1 and A2, indium (6$\%$) was added to tune the chemical potential.

\begin{figure*}[t]
\includegraphics[width=17 cm]{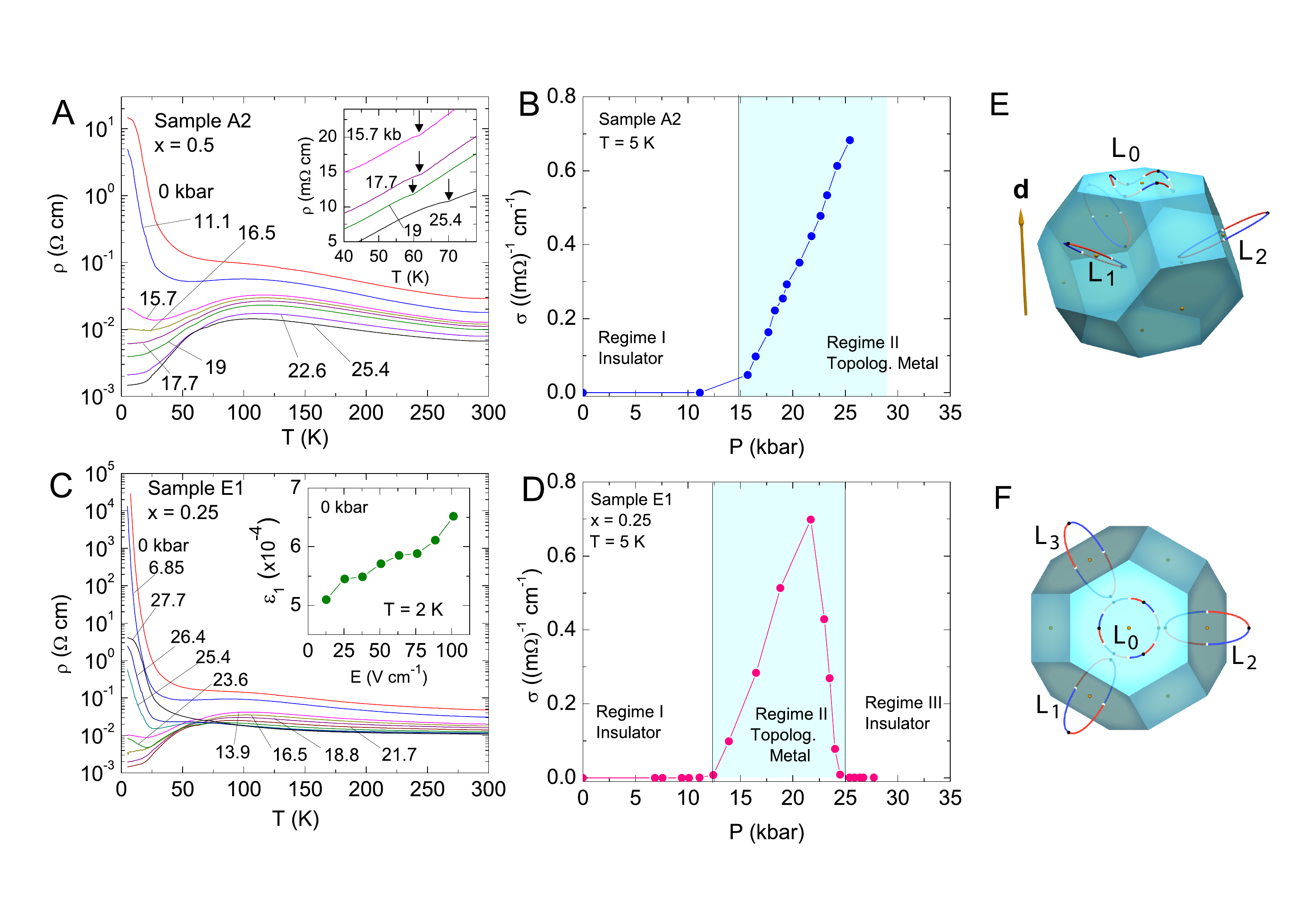}
\caption{\label{figRT} The phase diagram of Pb$_{1-x}$Sn$_x$Te inferred from the resistivity $\rho$ vs. temperature $T$ and pressure $P$.
Panel (A) plots curves of $\rho$ vs. $T$ in Sample A2 ($x = 0.5$) in zero $B$ measured at selected $P$. At 5 K, $\rho$ decreases by 4 orders of magnitude as $P\to$ 25.4 kbar (the IM transition). The inset shows the kinks (arrows) in $\rho$ (between 62 and 70 K) which signal a transition to a state with broken inversion symmetry. Panel (B) shows the steep increase of the conductivity $\sigma = 1/\rho$ at 5 K in the metallic phase (shaded in blue, $P>P_1$). Panel (C) shows $\rho$ vs. $T$ at selected $P$ in Sample E1 ($x$ = 0.25). The insulating state is recovered at $P_2\sim$ 25 kbar. The inset plots the dielectric response $\varepsilon_1$ measured vs. applied electric field $E$ at 2 K and ambient $P$ (a spontaneous value $\varepsilon_1\sim 5\times 10^4$ is measured as $E\to 0$). Panel (D) plots $\sigma$ vs. $P$ at 5 K to display the metallic state in E1 (shaded) sandwiched between $P_1$ and $P_2$. Panels (E) and (F) show the calculated Weyl node trajectories~\cite{SI}, magnified 10$\times$ relative to the Brillouin Zone (BZ) scale (with ${\bf B} = 0$). In (E), the vector ${\bf d}\parallel [111]$ (arrow) is the assumed ferroelectric displacement. Panel (F) shows the top view (sighting $\parallel\bf d$). Under pressure the 12 Weyl nodes at $L_1, L_2, L_3$ trace out elliptical orbits until they annihilate at the black points. The Weyl nodes at $L_0$ trace an orbit that undulates about a circular path (over a restricted pressure interval~\cite{SI}).
}
\end{figure*}

Figure \ref{figRT} provides an overview of the IM transitions in Samples A2 and E1. As $P$ increases from 0 (ambient) to 25.4 kbar, the resistivity profile, $\rho$ vs. $T$, changes from insulating to metallic behavior in A2 (Fig. \ref{figRT}A). Close examination reveals a kink in $\rho$ indicating a sharp transition at $T_c$ = 62 --70 K (arrows in inset). Figure \ref{figRT}B displays the rapid increase in the zero-$B$ conductivity $\sigma\equiv 1/\rho$ at 5 K as $P$ exceeds the critical value $P_1\sim$ 15 kbar for the IM transition. The resistivity curves in Sample E1 ($x$ = 0.25) are broadly similar except that $\rho$ at ambient $P$ attains much higher values at 5 K (3$\times 10^{4}\;\Omega$cm). As $P$ increases from ambient to $P_1$ (12 kbar) an IM transition occurs to a metallic state (with $\rho$ decreasing by over 7-decades at 5 K). In Samples E1 and Q1 (which have smaller Sn content than A1 and A2), the second critical pressure $P_2$ = 25 kbar is accessible in our experiment. The profile of $\sigma$ vs. $P$ at 5 K (Fig. \ref{figRT}D) displays the metallic phase sandwiched between the two insulating phases.

The end member SnTe is known to be ferroelectric~\cite{Brillson}, but the existence of ferroelectric distortion is less clear for finite Pb content. To establish inversion symmetry breaking, we performed dielectric measurements~\cite{SI} in Sample E1 which has a very large $\rho$ below 10 K ($>10^3$ $\Omega$cm). By varying the $E$-field (12$\to$100 V/cm), we show that a large spontaneous dielectric response $\varepsilon_1\sim 5\times 10^4$ exists in the limit $E\to 0$ (inset, Fig. \ref{figRT}C). The spontaneous polarization $\vec{\cal P}_s$ provides direct evidence that the insulating state below $P_1$ in E1 is ferroelectric. Although dielectric measurements cannot be performed in A2 (carrier screening is too strong), the kink in $\rho$ (arrow) implies that $\vec{\cal P}_s$ appears at 62 to 70 K.

In parallel, we performed \emph{ab initio} calculations~\cite{SI}, in which the lattice parameter $a$ is varied to simulate pressure. To break inversion symmetry, we assumed a weak ferroelectric displacement ${\bf d}\parallel$[111]. The calculations reveal that, above $P_1$, two pairs of Weyl nodes appear near each of the points $L_1$, $L_2$ and $L_3$ (equivalent in zero $B$; Figs. \ref{figRT}E and \ref{figRT}F). As $P$ increases, the 12 nodes trace out elliptical orbits (shown expanded 10$\times$ relative to the BZ), and eventually annihilate pair-wise at the black dots, consistent with the scenario in Refs.~\cite{MurakamiKuga,Okugawa}. The red and blue arcs refer to nodes with $\chi=1$ and -1, respectively. The splitting of the node at $L_0$ occurs in a much narrower pressure interval.

\begin{figure*}[t]
\includegraphics[width=15 cm]{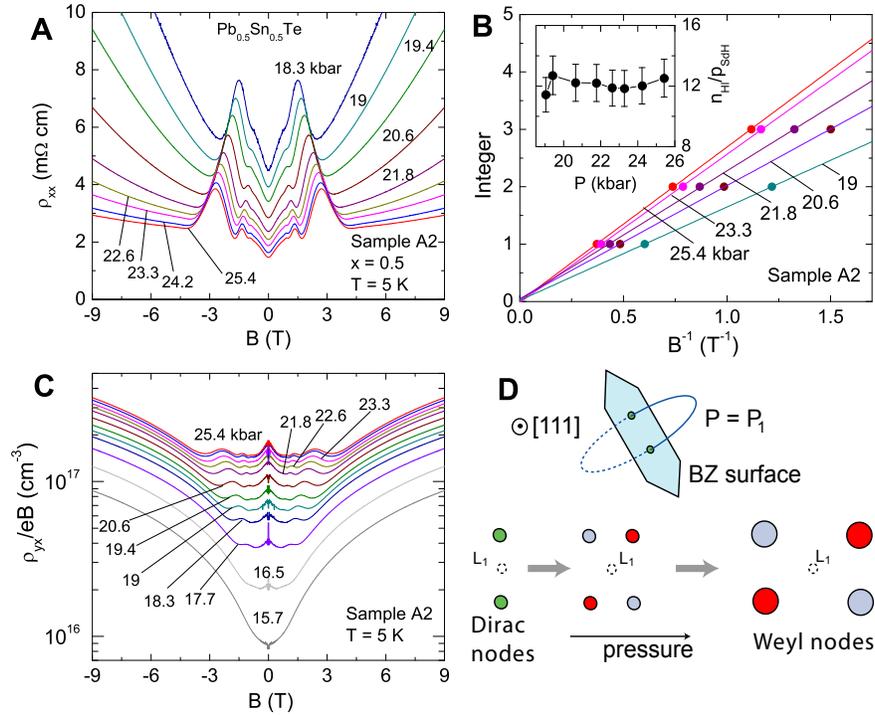}
\caption{\label{figSdH} 
The nucleation and growth of small FS pockets as $P$ exceeds $P_1$ in Sample A2, as observed by SdH oscillations. Panel (A) shows the resistivity $\rho_{xx}$ versus a transverse $\bf B$ measured at 5 K and with $P$ fixed at values 18.3 to 25.4 kbar. At each $P$, the oscillations below 3 T correspond to SdH oscillations (the largest peak corresponds to the $n$ = 1 LL). Panel (B) plots the inverse peak fields $1/B_n$ of $\rho_{xx}$ vs. the integers $n$. The slopes yield small extremal FS cross-sections $S_F$ (1.6-2.7 T) which increase with $\Delta P = P-P_1$. The inset shows that the ratio $n_H/p_{SdH}$ is 12$\pm 1$ independent of $P$ (see text). Panel (C) plots the Hall resistivity (divided by $Be$) $\rho_{yx}/Be$ vs. $B$. At low $B$ (where SdH oscillations occur), the flat profile allows $\rho_{yx}/Be$ to be identified with the total hole density $n_H$. The strong increase of $\rho_{yx}/Be$ above 3 T reflects the Berry-curvature term (Fig. \ref{figHallBerry}). Panel (D) shows a top-down view (along $[\bar{1}\bar{1}\bar{1}]$) of the $L_1$ hexagon face in zero $B$. At $P_1$, two Dirac nodes nucleate around $L_1$ because of inversion symmetry breaking. With increasing $P$, the 4 Weyl FS near $L_1$ move apart and expand in volume (chirality $\chi$ = 1 (red) and -1 (grey)). 
}
\end{figure*}

{\bf 
\vspace{3mm}\noindent
Quantum oscillations}\\
\noindent
The samples' high mobilities $\mu$ (20,000 to $4\times 10^6$ cm$^2$/Vs; Table in \cite{SI}) allow us to ``count'' the number of FS pockets by monitoring the Shubnikov de Haas (SdH) oscillations. As shown in Figs. \ref{figRT}B and D, $\sigma$ increases steeply vs. the reduced pressure $\Delta P =P-P_1$. Figure \ref{figSdH}A displays the resistivity $\rho_{xx}$ measured in Sample A2 in a transverse magnetic field $\bf B$ ($\parallel\bf\hat{z}$) at selected values of $P$. From the linear variation of $1/B_n$ vs. integers $n$ (with $B_n$ the peak fields in $\rho_{xx}$; see Fig. \ref{figSdH}B), we find that the FS caliper area $S_F$ increases from 1.6 to 2.7 T between 19 and 25.4 kbar. The tallest peak in Panel A corresponds to the $n$ = 1 Landau level (LL). The SdH-derived Fermi wavevector $k_F$ corresponds to a hole density $p_{SdH} = \frac43\pi k_F^3/(2\pi)^3$ per spin (assuming a spherical FS).

The sharp increase in hole density is also evident in the Hall resistivity $\rho_{yx}$ (which is $B$-linear in weak $B$). To highlight this, we plot the ratio $\rho_{yx}/Be$ vs. $B$ (Fig. \ref{figSdH}C). In weak $B$ (e.g. $|B|<$3.5 T in the top curve), the ratio is $B$-independent, which allows it to be identified with the Hall density $n_H$ (the abrupt increase above 3 T is the interesting anomalous Hall term discussed below). From $n_H$, we derive $\mu\sim$ 1.8$\times 10^4$ and 2.86$\times 10^4$ cm$^2$/Vs in A1 and A2, respectively, at 25 kbar. 

Crucially, we find that $n_H$ is always much larger than $p_{SdH}$, which implies a large number $N_F$ of identical pockets. The ratio $n_H/p_{SdH} = N_F$ equals 12$\pm 1$ over the whole pressure interval (inset, Fig. \ref{figSdH}B). Because a smaller $N_F$ (e.g. 4, 6 or 8) can be excluded, the results strongly support the choice ${\bf d}\parallel$[111] which creates 3 equivalent $L$ points.

\begin{figure*}[t]
\includegraphics[width=15 cm]{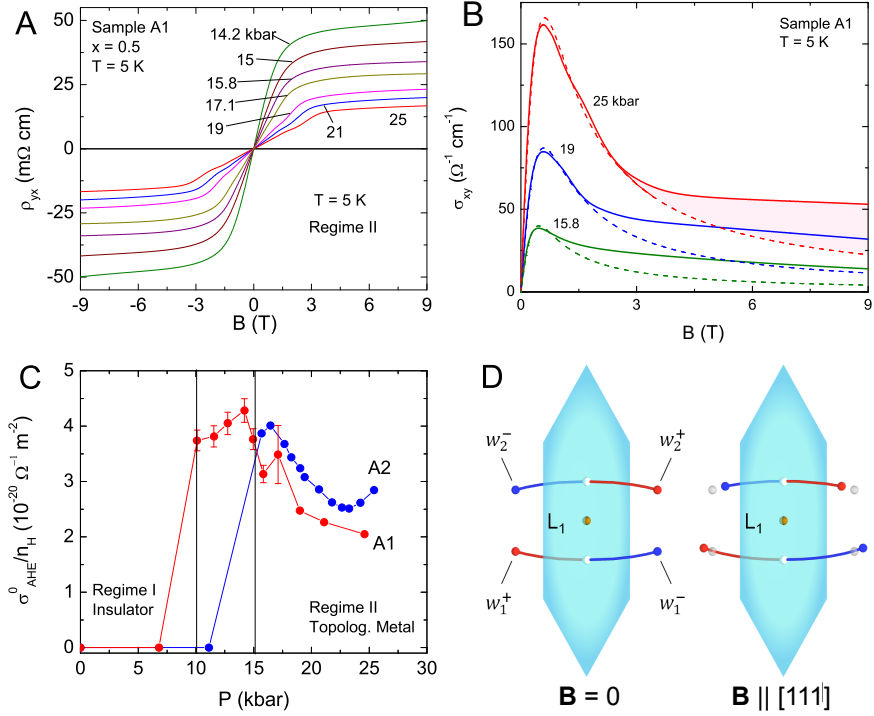}
\caption{\label{figHallBerry} 
The Berry curvature term in the Hall response. Panel (A) plots the observed curves of $\rho_{yx}$ vs. $B$ at 5 K with $P$ fixed at values above $P_1$. Instead of the conventional $B$-linear profile, $\rho_{yx}$ bends over at low $B$ (2-3 T), implying an extraordinary contribution to $\sigma_{xy}$ at large $B$. Panel (B) plots $\sigma_{xy}$ vs $B$ (derived from the measured $\rho_{ij}$) at 3 values of $P$. The Drude expression fits well to the curves at low $B$, but reveals the existence of an excess contribution (shaded in the curve at 25 kbar) at large $B$, identified with $\sigma^A_{xy}$ that increases with $B$. Panel (C) plots the ratio of $\sigma^0_{\mathrm{AHE}}/n_H$ (see fits in \cite{SI}) vs. $P$ in Samples A1 and A2 ($x = 0.5$). The ratio, proportional to $\langle\Omega_z\rangle$, shows a sharp increase at $P_1$ followed by a milder variation in the metallic phase. Panel (D) shows the effect of $\bf B$ on the Weyl node separations (viewed along $[\bar{1}\bar{1}\bar{1}]$). In zero $B$ (left sketch), the Weyl nodes are equal in size and symmetrically located about $L_1$ ($\mathbf{\Omega}$ vanishes). A finite Zeeman field (right) increases the separation of the pair $w_1^{\pm}$ while decreasing that of $w_2^{\pm}$. The explicit breaking of TRS leads to a finite $\sigma^A_{xy}$.
}
\end{figure*}

\vspace{3mm}\noindent 
{\bf Anomalous Hall Effect}\\
\noindent
We next describe the evidence for a topological metallic phase. The Hall resistivity $\rho_{yx}$ has a highly unusual field profile. As $B$ increases, the initial $B$-linear increase abruptly changes, bending over to a nominally flat profile (Fig. \ref{figHallBerry}A). At first glance, this recalls the anomalous Hall effect (AHE) in a ferromagnet~\cite{Nagaosa} where the intrinsic AHE arises from a large Berry curvature rendered finite by spontaneous breaking of TRS, but there is a subtle difference. In PbSnTe, TRS remains unbroken under $P$, so the AHE should be absent in the Weyl phase if $B = 0$ ($\bf \Omega$ cancels pairwise between Weyl nodes with $\chi = \pm 1$). However, when TRS is broken in finite $\bf B$, the cancellation is spoilt by the Zeeman energy (see below). The finite $\bf\Omega$ leads to a large AHE signal.

In weak-$B$ then (with $\Omega$ negligible), the initial slope of $\rho_{yx}$ is dominated by the ordinary Hall effect as evidenced by the linearity of $n_H$ vs. $p_{SdH}$ in Fig. \ref{figSdH}B (by contrast, in a ferromagnet the AHE term is dominant even in weak $B$, so the weak-$B$ $\rho_{yx}$ and $n_H$ are unrelated~\cite{Nagaosa}).

The total (observed) Hall conductivity $\sigma_{xy}$ is the sum of the conventional Hall and anomalous Hall conductivities, $\sigma^N_{xy}$ and $\sigma^A_{xy}$, respectively (Fig. \ref{figHallBerry}B). With $\sigma^N_{xy}$ given by the Drude expression, we find that a good fit to $\sigma_{xy}$ is achieved if we assume $\sigma^A_{xy} = \sigma^0_{\mathrm{AHE}}\,g(x)$ where $\sigma^0_{\mathrm{AHE}}$ is the AHE amplitude and $g(x)$ is the smoothed step-function $1/(\mathrm{e}^{-x}+1)$ with $x$ a reduced field~\cite{SI}. In terms of $\bf \Omega$, $\sigma^0_{\mathrm{AHE}}$ is given by~\cite{Nagaosa}
\be
\sigma^{0}_{\mathrm{AHE}} = \frac{e}{(2\pi)^3}\int d^3k \;\Omega_z({\bf k})\;f^0_{\bf k} = e\langle\Omega_z\rangle n_{tot}
\label{eq:sxy}
\ee
where $f^0({\bf k})$ is the Fermi-Dirac distribution, $\langle\Omega_z\rangle$ the Berry curvature averaged over the FS and $n_{tot}$ the total carrier density. From the fits at each $P$, we can track the variation of $\langle\Omega_z\rangle \sim \sigma^{0}_{\mathrm{AHE}}/n_{tot}$ vs. $P$. As shown in Fig. \ref{figHallBerry}C, the curvature $\langle\Omega_z\rangle$ is negligible below $P_1$, but becomes large and finite in the metallic phase consistent with the Weyl scenario.

In Fig. \ref{figHallBerry}B, we plot the observed $\sigma_{xy}$ (solid curves) together with the Drude curve for $\sigma_{xy}^N$ (dashed curves). Their difference is $\sigma^A_{xy}$ (shaded region in the curve at 25 kbar). Similar results are obtained in A2 and E1~\cite{SI}.
Interestingly, $\sigma^A_{xy}$ grows quite abruptly at an onset field $B_A$ close to where the system enters the lowest ($n=0$) LL. Above $B_A$, the increasing dominance of the AHE current accounts for the abrupt bending of $\rho_{yx}$ already noted in Fig. \ref{figHallBerry}A, as well as the sharp increase above $B_A$ in $\rho_{yx}/Be$ in Fig. \ref{figSdH}C. The observation that $\sigma_{xy}^A$ is most prominent within the $n=0$ LL (which is strictly chiral for Weyl fermions) suggests to us that it is intimately related to the chirality of the nodes.

Each Weyl node acts as a source ($\chi$ = 1) or sink (-1) of $\mathbf{\Omega}$. As mentioned, in zero $B$, TRS requires the net sum of $\mathbf{\Omega}$ over each pair of Weyl nodes to vanish (Fig. \ref{figHallBerry}D). The \emph{ab initio} calculations~\cite{SI} reveal how this cancellation is spoilt when TRS is broken in finite $\bf B$. A finite Zeeman field $\lambda$ shifts the band energies, depending on their spin texture. This increases the $\bf k$-space separation and Fermi energy of one pair, say $w_1^{\pm}$, while reducing them in $w_2^{\pm}$ (Fig. \ref{figHallBerry}D). The unbalancing creates a finite $\Omega\sim \sigma^{A}_{xy}$ that grows with $B$~\cite{SI}. 

\begin{figure*}[t]
\includegraphics[width=15 cm]{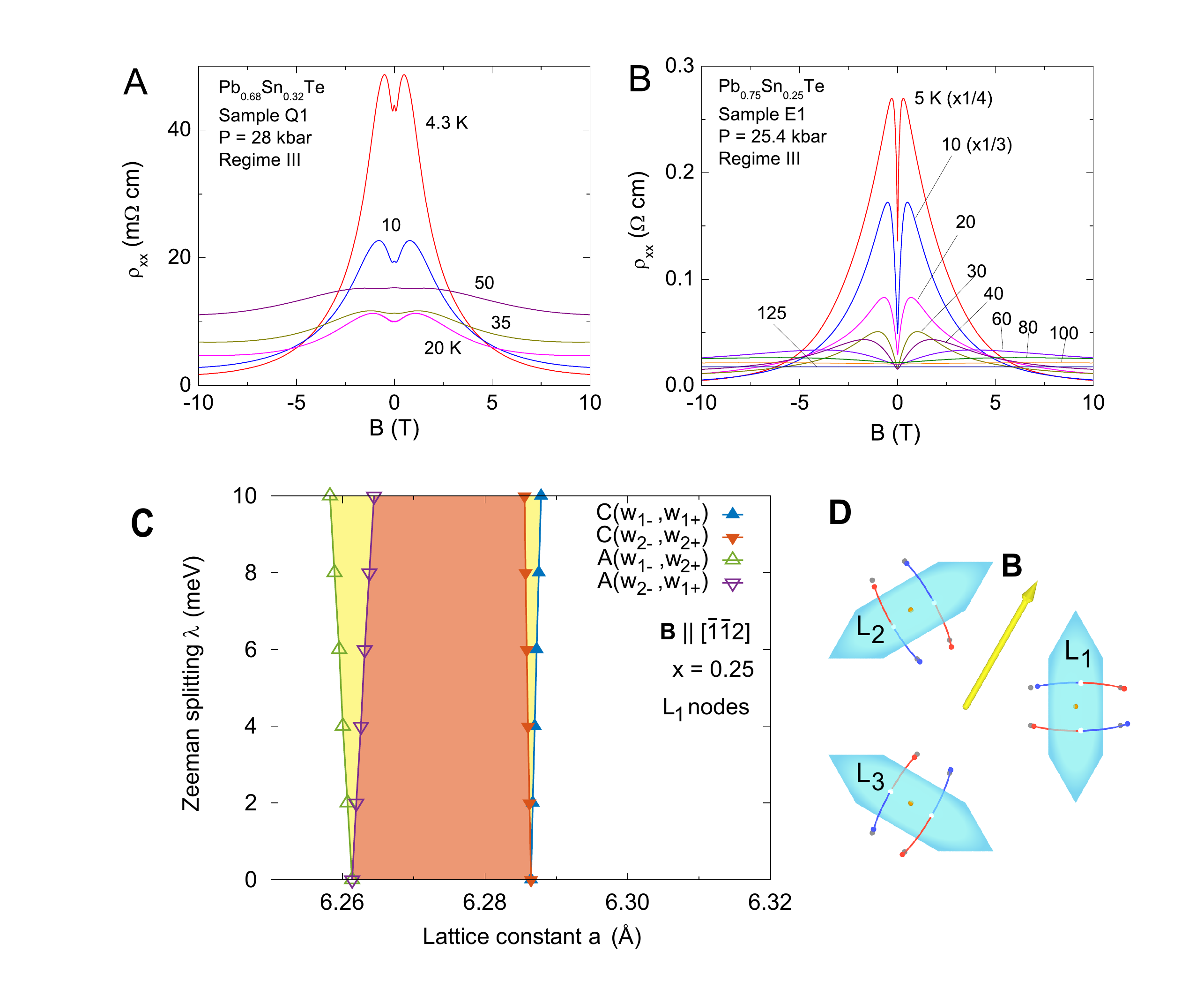}
\caption{\label{figMR} 
Large negative magnetoresistance (MR) observed when $P$ is fixed just above $P_2$ in Samples Q1 and E1. Panel (A) plots $\rho_{xx}$ vs. $B$ in Q1 ($x = 0.32$) at selected $T$ with $P$ = 28 kbar. At $P>P_2$, both samples are insulators at $B$ = 0. However, at $B$ = 10 T, both are metallic ($\rho_{xx}$ decreases with $T$). In Panel (B), similar curves show an even larger negative MR in Sample E1 ($x = 0.25$). Panel (C) shows the \emph{ab initio} phase diagram~\cite{SI} in the plane of $a$ vs. $\lambda$ at $L_1$ for $\bf B\parallel[\bar{1}\bar{1}2]$. With increasing $\lambda$, the Weyl-node annihilation boundary (left $V$-shaped wedge shaded yellow) expands. A weaker expansion occurs at creation. Panel (D) shows the effect of $\bf B$ (arrow) on the locations of the Weyl nodes around $L_1$, $L_2$ and $L_3$.
}
\end{figure*}

\vspace{3mm}\noindent 
{\bf Giant Magnetoresistance}\\
\noindent
Perhaps the most dramatic feature in Pb$_{1-x}$Sn$_x$Te is the appearance of giant negative magnetoresistance (MR) at pressures just above $P_2$. 
In Fig. \ref{figMR}A, we display the MR curves in Sample Q1 for selected $T$ with $P$ fixed at 28 kbar ($\sim$3 kbar above $P_2$). At 4.3 K, $\rho_{xx}$ decreases by a factor of 30 as $B$ increases to 10 T (aside from a slight dip feature below 0.5 T). In Fig. \ref{figMR}B, similar curves for E1 (with $P$ = 25.4 kbar) display an even larger negative MR (the weak-$B$ dip feature is more prominent as well). The large negative MR is steadily suppressed as we increase $P$ away from the $P_2$ boundary. The negative MR magnitude is similar in magnitude in both the transverse MR and longitudinal MR geometries ($\bf B\parallel z$ and $\bf B\parallel x$, respectively). This implies a Zeeman spin mechanism. Finally, we note that, in both Q1 and E1, $\rho_{xx}$ measured at 10 T decreases as $T\to$ 5 K (metallic). 

\vspace{3mm}\noindent 
{\bf Discussion}\\
\noindent
The anomalously large changes in $\rho$ imply that the insulating state (at zero $B$) is converted to a metallic state in finite $B$. This is confirmed in the calculation~\cite{SI}. A large $\lambda$ favors the Weyl phase (the left $V$-shaped yellow region in Fig. \ref{figMR}C). As the phase boundary now tilts into the insulating side, the metallic phase is re-entrant in increasing $B$. The observation of the giant negative MR provides further evidence in support of the Weyl node scenario.

As predicted in Refs.~\cite{Murakami1,MurakamiKuga,Okugawa,YangNagaosa}, gap closing in materials lacking inversion symmetry leads to a metallic phase that is protected by the distinct chirality of Weyl nodes. Pb$_{1-x}$Sn$_x$Te is an instructive first example. Increasing $P$ drives an IM transition at $P_1$, with $\rho$ (at 5 K) falling by 4 to 7 orders of magnitude. Above $P_1$, the growth of the FS calipers is tracked by large SdH oscillations. The number of nodes ($\sim 12$) is consistent with the appearance of 4 Weyl nodes at each of the 3 $L_1$ points. The Berry curvature $\mathbf{\Omega}$, rendered finite in $B$, leads to an anomalous Hall effect that is most prominent in the $n=0$ Landau level. Finally, we find that the boundary $P_2$ is shifted in finite $B$. The re-entrance of the metallic phase leads to a dramatic decrease in $\rho_{xx}$ by a factor of 30-50. 

\vspace{3mm}\noindent 
{\bf Materials and Methods}\\
\noindent
{\bf Crystal growth}\\
\noindent
Single crystals of Pb$_{1-x}$Sn$_x$Te were grown by the conventional vertical Bridgman technique. High purity elements (5N) with the targetted values of $x$ were sealed in carbon coated quartz tubes under a high vacuum of $\sim 10^{-5}$ mbar. The ampoules were heated at 1050 C for 12 hrs. To insure homogeneous mixing of the melt and to avoid bubble formation in the bottom, the ampoules were stirred. The ampoules were slowly lowered through the crystallization zone of the furnace, at the rate of 1 mm/h. High quality single crystals boules of length $\sim$10 cm were obtained. The crystal boules were cut into segments of 1 cm to investigate the bulk electronic properties along the boule length. The crystals were easily cleavable along different crystallographic planes.

A major impediment in the rocksalts is to ensure that the chemical potential of the alloy lies within the bulk gap (otherwise the pressure induced changes to the gap will not be observable). 
To achieve this goal in crystals with Sn content $x$ = 0.5, we have found it expedient to dope the starting material with indium (at the 6$\%$ level, with composition (Pb$_{0.5}$Sn$_{0.5}$)$_{1-y}$In$_y$Te, with $y$ = 0.06. Indium doping has previously been carried out and investigated by several groups to understand the superconducting phase in Pb$_{1-x}$Sn$_x$Te~\cite{Novak,Cava,Zhong}. Zhong \etal~\cite{Zhong} have reported that In-doped Pb$_{1-x}$Sn$_x$Te ($x = 0.5$) induces an insulating behavior. However, in our judgment, the precise role of In doping in the Pb-based rocksalts is not well understood, and merits further detailed investigation.

The X-ray diffractograms recorded for two powdered specimens of two typical samples are shown in Fig. S4 of Ref.~\cite{SI}. The grown crystals are single-phased. The diffraction peaks are in very close agreement with the rocksalt crystal structure of space group $Fm\bar{3}m$.

\begin{table}
\begin{tabular}{|c|c|c|c|c|c|} \hline
Sample	&   $x$				&  $\mu$ 	                     & $n_H$                           &  $\sigma$                        &   $P$       \\
            &                &   (cm$^2$/Vs)             &   (cm$^{-3}$)	             &  (m$\Omega$ cm)$^{-1}$&  kbar       \\ \hline\hline
A1    &    0.5         &    18,000		                 &   1.59$\times 10^{17}$ &   0.41										        &  25         \\ \hline
A2   &    0.5         &    29,000                    &   1.56$\times 10^{17}$  &   0.68                           & 25.4     \\ \hline
E1   &    0.25       &   500,000                   &   9.35$\times 10^{15}$  &   0.70			                     & 21.7			\\ \hline
Q1  &    0.32       &   4.2$\times 10^6$    &   -1.05$\times 10^{16}$  &   7.02		                     & 21 					\\ \hline
\end{tabular}
\caption{\label{Tab1}
Parameters of samples of Pb$_{1-x}$Sn$_x$Te investigated. Columns 2, 3, 4 and 5 report the Sn content $x$, the mobility $\mu$, the
Hall carrier density $n_H$ and the conductivity $\sigma$ (at $B=0$), respectively. The minus sign in $n_H$ (Sample Q1) indicates $n$-type carriers.
All quantities in the table were measured at 5 K at the pressure $P$ given in the last column. Samples A1 and A2 are slightly doped with In to tune the chemical potential (composition (Pb$_{0.5}$Sn$_{0.5}$)$_{1-y}$In$_y$Te, with the In content $y$ = 0.06).
}
\end{table}

Some of the parameters measured in the 4 samples investigated (A1, A2, E1 and Q1) are reported in Table \ref{Tab1}. The mobilities in E1 ($p$ type) and Q1 ($n$-type) are very high (500,000 and 4.2$\times 10^6$ cm$^2$/Vs, respectively). Although the samples with $x$ = 0.50 (A1 and A2) have lower mobilities (18,000 and 29,000 cm$^2$/Vs, respectively), clear SdH oscillations are observed above $P_1$.

\begin{figure*}[t]
\includegraphics[width=15 cm]{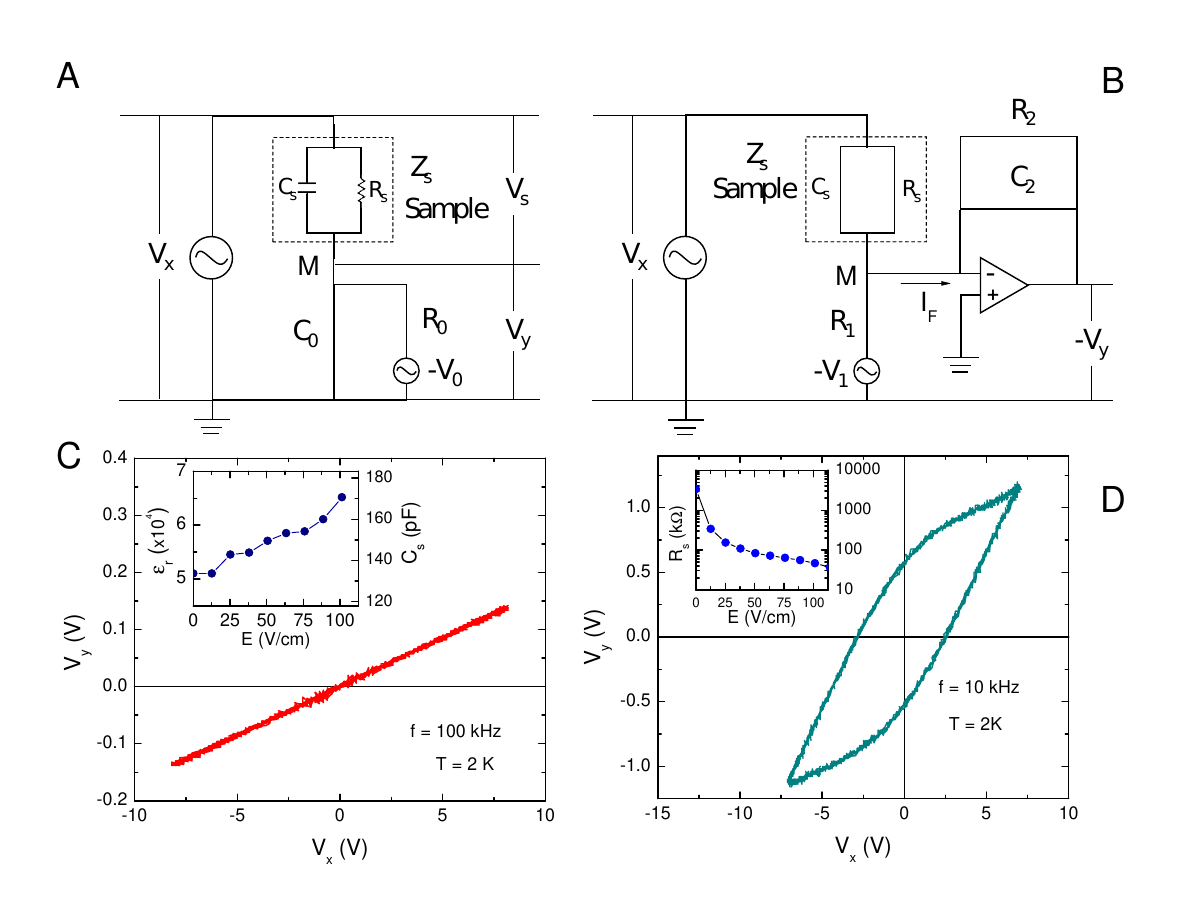}
\caption{\label{ST} 
Panel A shows the electrical circuit for measuring the relative dielectric constant $\varepsilon_r$ of the sample. The voltage $V_y \sim V_x~C_s/C_0 \ll V_x$ is proportional to the dielectric displacement $D$. By contrast, the applied voltage $V_x$ is proportional to the electrical field $E$ across the sample. Therefore, the curve of $V_x$ versus $V_y$ gives the relation between $E$ and $D$, yielding the relative dielectric constant $\varepsilon_r$ as a function of applied electric field $E$. Panel B shows the modified electrical circuit of Panel A. The op-amp is introduced such that the point M is better virtually grounded. The integration circuit gives $V_y = V_x~C_s/C_2$ when $V_1/R_1 = V_x/R_s$ is satisfied, again proportional to $D$. Since $C_2$ does not have to satisfy $C_2\gg C_s$ for Panel B, large measured signal $V_y$ can be obtained. Panel C shows the relation between $E$ and $D$ ($V_x$ v.s. $V_y$), yielding relative dielectric constants as large as $\varepsilon_r \sim 5\times 10^4$ even in the limit of $E \rightarrow 0$ as shown in the inset, suggesting the system is in the FE state. Panel D shows the apparent nonlinear relation between $V_x$ v.s. $V_y$ due to the nonlinearity of the resistive component $R_s$ of the sample. The inset shows how $R_s$ varies with increasing $E$.
}
\end{figure*}

\vspace{3mm}\noindent 
{\bf Measurement of dielectric constant}\\
\noindent
In this section, the experimental setup for the measurements of the relative dielectric constants $\varepsilon_r$ (hence the capacitive component of the sample impedance) is summarized. For this purpose the (modified) Sawyer-Tower method is employed~\cite{Sawyer}. The circuit of the original Sawyer-Tower method is shown in Panel A in Fig.~\ref{ST}. Panel B shows the modified method using an operational amplifier (op-amp)~\cite{Yamaguchi}.

For the setup shown in Panel A, the sample with a capacitance component $C_s$ and a resistance component $R_s$ is connected in series with a known reference capacitance $C_0\gg C_s$. This reference capacitance $C_0$ is connected in parallel with a fixed resistor $R_0$ in series with an adjustable voltage source $V_0$ (or equivalently a variable resistor $R_c$) that is set such that voltage $V_y$ is in-phase with voltage $V_x$, where $V_x$, $V_y$ represent respectively the voltages across the whole electrical circuit and the reference capacitor $C_0$. Since $C_0\gg C_s$ is satisfied, $V_s \sim V_x$, $V_y \sim V_x~C_s/C_0 \ll V_x$ hold, where $V_s$ is the voltage applied aross the sample. This means that the point M in the figure can be treated as the virtual ground, which immediately yields $V_0/R_0 \sim V_x/R_s$.

The setup in Panel B is the same as Panel A except that now the op-amp is directly used to provide a more stable virtual ground at point M. From point M, current $I_F$ is driven to the op-amp which is connected to $C_2$ and $R_2$ in parallel. Here, a large resistor $R_2$ ($R_2C_2\gg 1/\omega$) is connected in parallel as the leak resistor so that the capacitor $C_2$ does not overload. This is the integration circuit so it gives $V_y = 1/C_2\int I_F dt$. If the adjustable voltage source $V_1$ is set such that it cancels the current $I_s$ flowing into the resistive component of the sample, i.e., if $V_1/R_1 = V_x/R_s$ is satisfied, the current $I_F$ equals the current flowing into the capacitive component of the sample $I_s = dQ/dt = d(C_sV_x)/dt$. This gives $V_y =  V_x~ C_s/C_2$ similar to the expression obtained for the case of Panel A. The advantage of using the op-amp is that one can choose any value of $C_2$ (as long as it satisfies $R_2C_2\gg 1/\omega$) so that one can get a larger measured signal $V_y$ than the case in Panel A.

In general, the two setups in Panels A and B work well. However, because the working frequency range of our op-amp (LF356N) is between 30~Hz and 30~kHz, we used setup A above 30~kHz. For frequencies below 30~kHz, both setups A and B were used. We confirmed that the two setups yield the same results.

The measured results of the relative dielectric constants are plotted as Panel C for $f$ = 100~kHz and Panel D for $f$ = 10~kHz in Fig.~\ref{ST}. Panel C shows that $V_y =  V_x~ C_s/C_2 = DS/C_2$ is proportional to $V_x = E~t$, with $E$ the applied electric field across the sample, $D$ the dielectric displacement, $S$, $t$ the area and the thickness of the sample respectively. Putting the parameters of the dimensions $S = 0.234$~mm$^2$, $t = 0.79$~mm, yields the relative dielectric constants $\varepsilon_r \sim 5\times 10^4$ plotted as the inset of Panel C in Fig.~\ref{ST}. Unfortunately, the sample undergoes dielectric breakdown above $E \sim 100$~V/cm, preventing us from observing the saturation of dielectric displacement $D$ at higher electric fields as should be expected from the ferroelectric (FE) behavior. Such kind of large relative dielectric constants $\varepsilon_r \sim 5\times 10^4$ even in the limit of $E\rightarrow 0$, however, strongly suggest that the system is in the FE state.

Here we remark that the nonlinear behavior shown in Panel D of Fig.~\ref{ST}, previously interpreted in Ref.~\cite{Klaus} as evidence for the FE state, is actually \textit{not} the manifestation of the FE state, but arises instead from the fact that the resistive component of the sample $R_s$ is strongly $E$-dependent as shown in the inset. Since $R_s$ is nonlinear, it is not possible to compensate the current $I_s$ flowing through the resistive component $R_s$, because $V_1/R_1 = V_x/R_s$ cannot be satisfied for every $V_x$ unless $V_1$ is changed nonlinearly. Therefore, if $V_1$ is set to compensate the $R_s$ at some fixed value of $V_x$ (7~V in the case of Panel D), then other parts of $V_x$ cannot be compensated. As a result, they produce a nonlinear behavior that looks like the saturation of FE state. The way to avoid this is to use the higher frequency $f$ so that a larger portion of the current flows into the capacitive component $C_s$ rather than the resistive component $R_s$. This is exactly the case shown in Panel C ($f$ = 100~kHz).


\vspace{1cm}\noindent
{\bf Acknowledgements}\\
The experimental project is supported by the US Army Research Office (W911NF-16-1-0116) and by the Gordon and Betty Moore Foundation’s EPiQS Initiative through Grant GBMF4539 (to N.P.O.). R.J.C. acknowledges an NSF-MRSEC grant DMR 1420541 (crystal growth). N. K. is supported by NSF-Partnership in Research and Education in Materials (PREM) Grant DMR-1205734 (calculations). 

\vspace{3mm}
\noindent
{\bf Author contributions:} T.L. conceived the idea of applying pressure to PbSnTe, and developed the experimental program with N.P.O. The samples were grown by S.K., Q.G. and R.J.C. The measurements were performed by T.L. with some assistance from J.L. Analyses of the results were performed by T.L., N.P.O., J.W.K. and N.K. The ab initio calculations were performed by J.W.K. and N.K. The manuscript was written by T.L. and N.P.O. with inputs from J.W.K. and N.K.  

\vspace{3mm}
\noindent
{\bf Competing interests:} 
The authors declare that they have no competing interests. 

\vspace{3mm}
\noindent
{\bf Data and materials availability:} 
The data are available from T.L. (liang16@stanford.edu) or N.P.O. (npo@princeton.edu)


\clearpage
\newpage

\renewcommand{\thefigure}{S\arabic{figure}}
\renewcommand{\thesection}{S\arabic{section}}
\renewcommand{\theequation}{S\arabic{equation}}

\setcounter{equation}{0}
\setcounter{figure}{0}
\setcounter{table}{0}
\setcounter{section}{0}

\vspace{4mm}
{\bf Supplementary Information\\
} 
\vspace{6mm} 

\section{\emph{Ab initio} Band calculations}

We have performed extensive density-functional theory calculations of the electronic structure of Pb$_{1-x}$Sn$_x$Te to investigate how changes to the lattice parameter affect the states in the vicinity of the $L$ points of the Brillouin zone (BZ) surface (this simulates application of hydrostatic pressure $P$). We assume the existence of a weak sublattice distortion ${\bf d}\parallel [111]$. At a specific pressure (identified as $P_1$), we find that a pair of Dirac nodes appears near each of the 3 equivalent $L$ points. As $P$ is raised above $P_1$, each Dirac node splits into a pair of Weyl nodes with opposite chirality. The Weyl nodes move in an elliptical orbit until they mutually annihilate at a higher pressure (identified as $P_2$). We also simulate the effect of time-reversal symmetry breaking on the Weyl pairs by introducing a finite Zeeman field. In general, the calculations agree well with the transport results reported in the main text.

The \textit{ab initio} calculations were carried out within the framework of the projector augmented-wave formalism [24], as implemented in the Vienna \textit{ab initio} simulation package (VASP) [25].
For the total-energy calculations, the PBEsol [26] was employed to treat the exchange-correlation potential. On the other hand, to investigate the band-gap evolution versus $P$, we employed the modified Becke-Johnson Local Density Approximation (MBJLDA) functional, which has been shown [27] to yield accurate
band gaps, effective masses, and  frontier-band ordering. Accurate determination of these parameters are especially important in the topological insulator phase. Spin-orbit coupling was included in the self-consistent calculations.

The energy cut-off of the plane-wave expansion of the basis functions was set at 300 eV and an $8 \times 8 \times 8$ $k$-point mesh was used in the Brillouin zone sampling. The calculated equilibrium lattice constants for SnTe and PbTe of 6.288 {\rm\AA} and 6.441 {\rm\AA}, respectively, are in good agreement with the experimental values of 6.303 {\rm\AA} and 6.443 {\rm\AA} [28].

We employ the WANNIER90 package [29] with a frozen energy window of  1 eV above the Fermi level $E_F$ to construct Wannier functions from the outputs of the first-principles calculations for the SnTe and PbTe pristine materials. In order to determine the electronic structure of the Pb$_{1-x}$Sn$_x$Te alloys we employ the virtual-crystal approximation where each Pb or Sn is replaced by a ``virtual" atom whose properties are the weighted average of the two
constituents. Further, for the lattice constant and the relative ferroelectric displacement $x_P$ sampling mesh, we use a linear interpolation scheme.

Even though the MBJLDA functional gives band gap of 0.24 eV and -0.13 eV for PbTe and SnTe, respectively, in reasonable agreement with the corresponding experimental values of 0.19 eV and  -0.20 eV [30],
the larger calculated values lead to large values of the predicted critical pressures $P_1$ and $P_2$. Therefore, using scissor operators the on site energies of the Pb and Sn $p$-orbitals are adjusted  to
reproduce the experimental values of the band gap at the theoretical equilibrium lattice for PbTe and SnTe, namely,
\begin{equation}
	\begin{split}
		\epsilon_{\text{Pb-}p} &\longrightarrow \epsilon_{\text{Pb-}p} - 0.047 \text{ eV} \\
		\epsilon_{\text{Sn-}p} &\longrightarrow \epsilon_{\text{Sn-}p} - 0.124 \text{ eV}.
	\end{split}
\end{equation}

The response of the Weyl nodes to a magnetic field $\bf B$ is simulated by including the Zeeman energy. The direction of $\bf B$ is determined by the spinor quantization axis in the VASP calculations
\begin{equation}
	H_{m,n}^{\text{Zeeman}}(\mathbf{k})=\lambda \sigma_{\hat{n}}.
\end{equation}
We investigated the two field directions ${\bf B} \parallel [111] $ and ${\bf B}\parallel [\bar{1}\bar{1}2]$. The evolution of the Weyl nodes under hydrostatic pressure and/or magnetic field is determined using a steepest descent method.

\subsection{Emergence of Weyl phase under pressure}
\begin{figure}[h!]
	\centering
	\includegraphics[width=9cm]{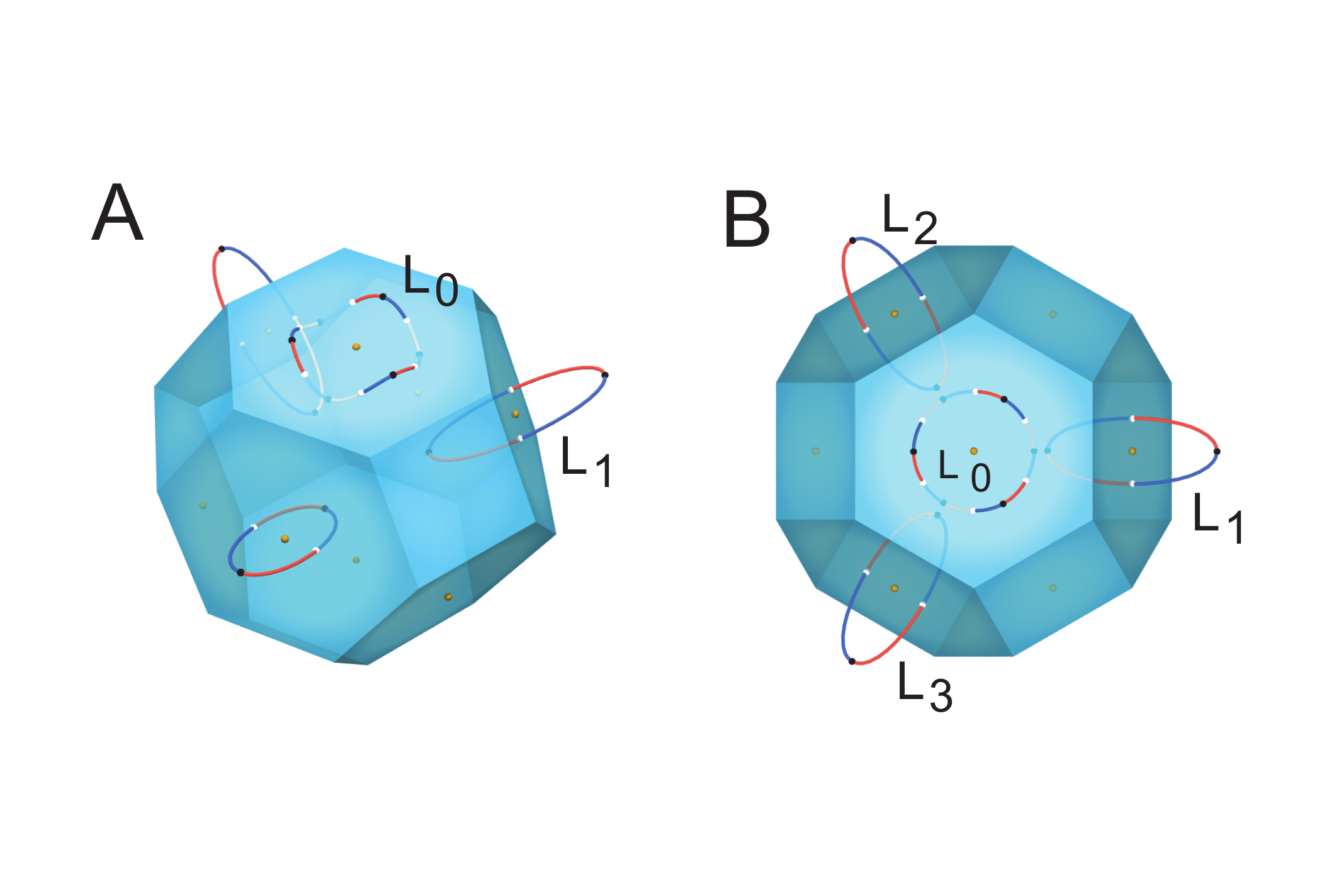}
	\caption{The calculated $\bf k$-space trajectories of Weyl nodes in Pb$_{1-x}$Sn$_x$Te ($x$ = 0.5) under applied pressure in zero magnetic field. Panel (A) shows the $L_0$ point and the 3 equivalent $L_1$ points on the BZ surface in a perspective with $\bf d\parallel$[111] vertical. Panel (B) shows the top-down view sighted along $\bf d$. Trajectories of Weyl nodes with positive (negative) chirality are colored red (blue). White and black dots indicate creation and annihilation points, respectively. The orbits are shown magnified by a factor of 10 relative to the scale of the BZ.
	}
	\label{fig:Weyl}
\end{figure}

In the absence of the ferroelectric displacement, band inversion under pressure occurs at the four equivalent $L$ points in $\bf k$ space. The ferroelectric atomic displacement $\bf d\parallel$ [111], which is observed in SnTe and GeTe, introduces a Rashba splitting that breaks the cubic symmetry, rendering the $L_0$ point (which lies in the direction [111]) inequivalent to the remaining three $L$ points (in zero $B$, these equivalent points are called collectively $L_1$; see Fig. \ref{fig:Weyl}). At the insulator-to-metal transition (at the lower critical pressure $P_1$), closing of the bulk gap at $L_1$ coincides with the appearance of two Dirac nodes bracketing each $L_1$ point. Increasing $P$ slightly above $P_1$ further splits each Dirac node into a pair of Weyl nodes of opposite chirality (there are 6 Weyl pairs altogether).


When the pressure exceeds $P_1$, the two Dirac nodes at $L_1$ split into four Weyl nodes, $w_{1}^{\pm}$ and $w_{2}^{\pm}$, where subscripts (1, 2) identify the starting Dirac node and superscripts (+, -) refer to the chirality of the Weyl nodes. In Fig. \ref{fig:Weyl}, the $\bf k$-space trajectories of the nodes with +(-) chirality are colored red (blue). For clarity, the orbits are displayed magnified by a factor of 10 relative to the BZ scale. White and black dots indicate the creation and annihilation points, respectively. As noted, we have 12 Weyl nodes altogether. As $P$ exceeds $P_1$, the nodes $w_1^-$ and $w_2^+$ move into the interior of the first BZ, whereas $w_1^+$ and $w_2^-$ move away from the interior. When $P$ reaches the higher critical pressure $P_{2}$, the 12 Weyl nodes annihilate pair-wise simultaneously (in zero $B$) at the points shown as black dots. Near the $L_0$ point, the Weyl nodes exist over a much narrower interval of $P$ between $P_1$ and $P_2$. The undulating trajectory of the Weyl nodes near $L_0$ under pressure (Fig. \ref{fig:Weyl}) is similar to that reported in ferroelectric hexagonal BiTeI [31].

\begin{figure}[h!]
	\centering
	\includegraphics[width=9cm]{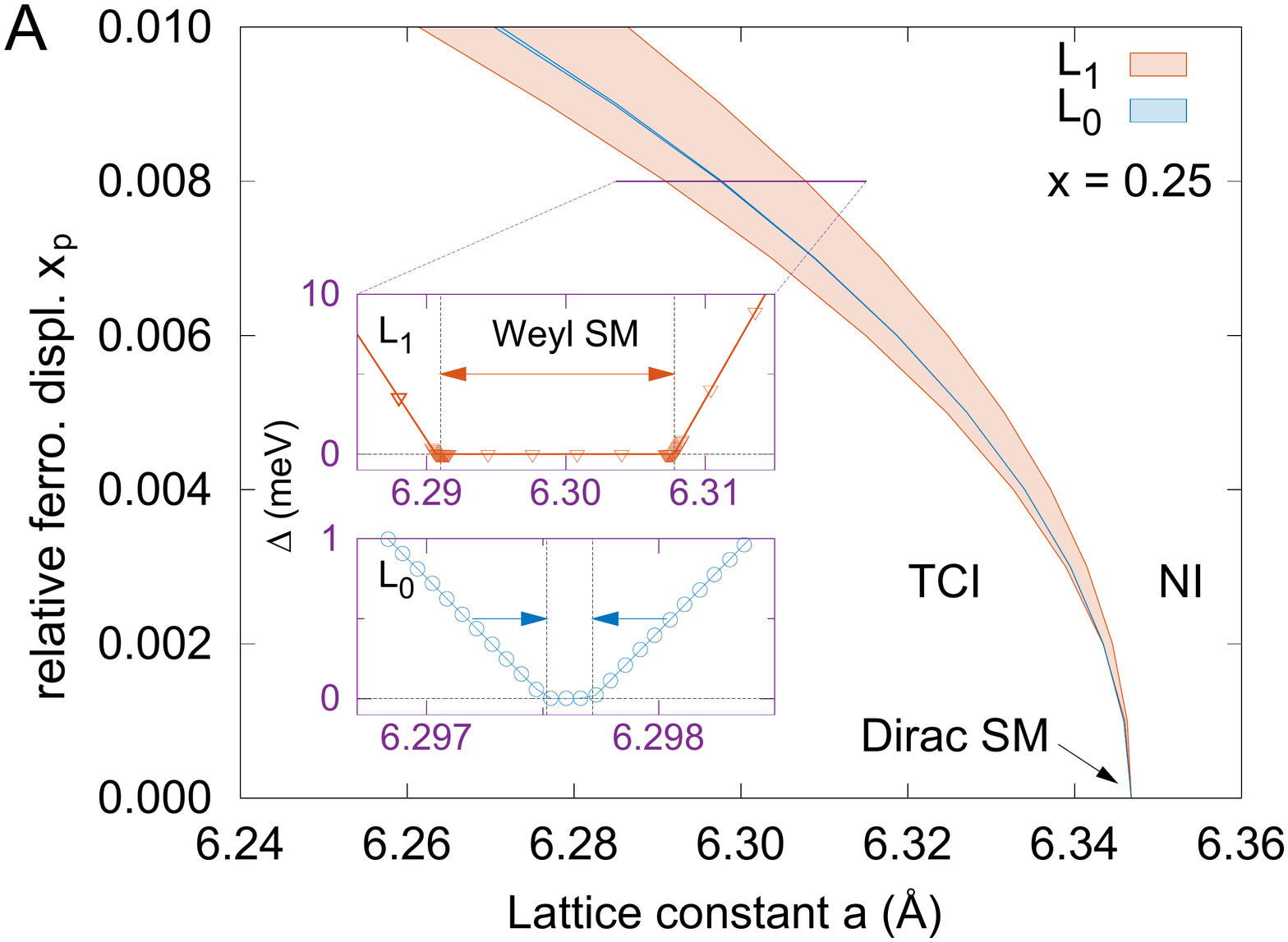}
	\includegraphics[width=9cm]{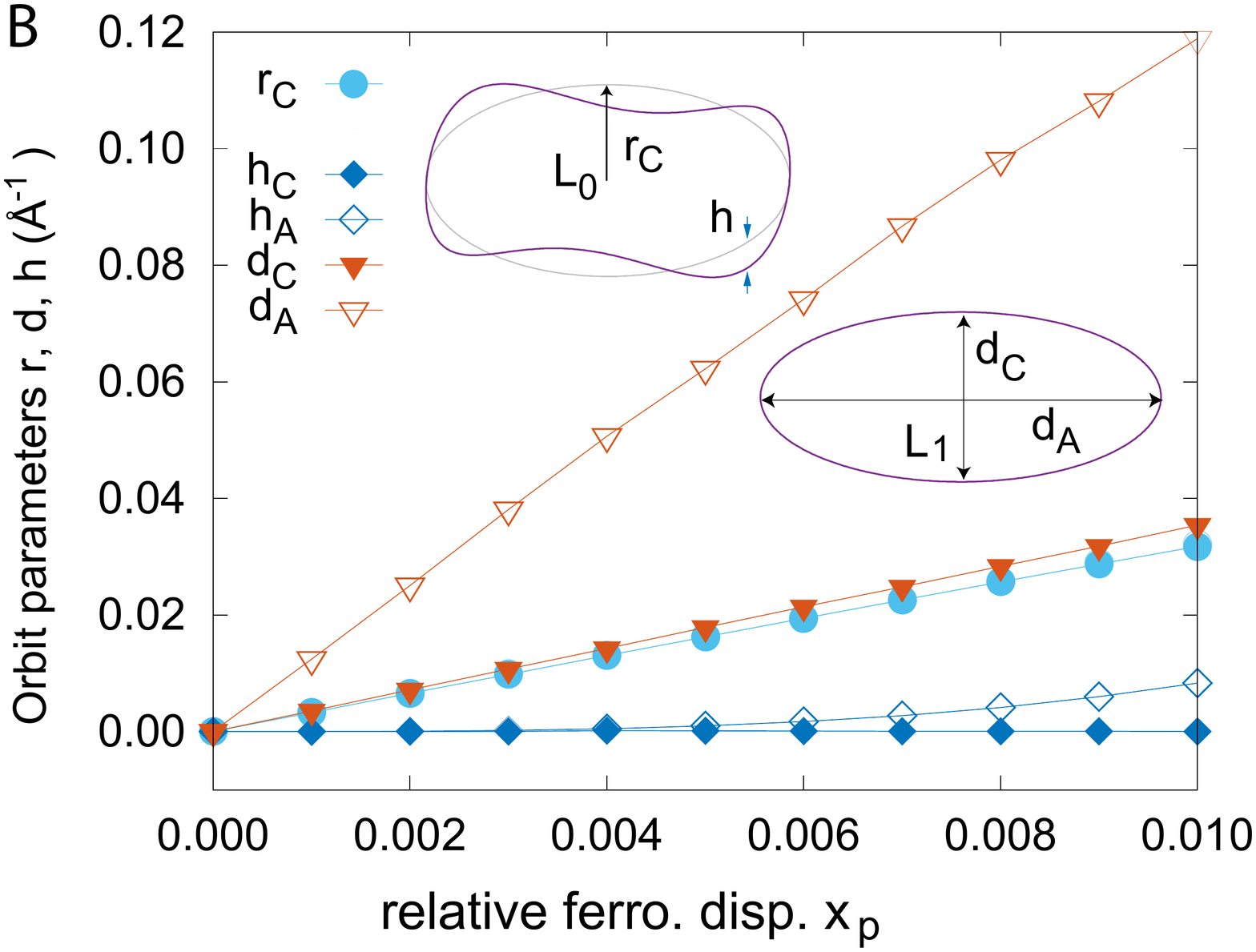}
	\caption{Panel (A): The phase diagram of the Weyl phase in Pb$_{1-x}$Sn$_{x}$Te ($x=0.25$) in the plane of $a$ vs. $x_p$ where $a$ is the lattice constant and $x_p$ the relative ferroelectric displacement ($x_p\equiv d/d_{[111]}$ with $d_{[111]}$ the unit-cell diagonal along [111]). The pink wedge is where the 12 Weyl nodes near the points $L_1$, $L_2$ and $L_3$ are stable. The narrow blue strip indicates where the Weyl nodes exist around the $L_0$ point. The wedge separates the topological crystalline insulator (TCI) and normal insulator (NI) phases. The upper inset shows how the bulk gap $\Delta$ at $L_1$ vanishes within the lattice interval $a$ = (6.292, 6.308) ${\rm\AA}$, with $x_p$ fixed at 0.008. The lower inset shows the bulk gap at $L_0$ closing within a much narrower interval of $a$.
		Panel (B) plots the increase versus $x_p$ of $r_C$ (the radius of the circular orbit of the Weyl nodes around $L_0$), and $d_A$ and $d_C$ (the major and minor diameters of the elliptical orbit around $L_1$ as sketched in the insets). Subscripts A and C refer to annihilation and creation. Also plotted is the undulation amplitude $h$ of the orbit at $L_0$. 
	}
	\label{fig:phase}
\end{figure}

In Fig. \ref{fig:phase}A, we plot the phase diagram in the plane of $a$ vs. $x_p$ to identify the Weyl phase in  Pb$_{1-x}$Sn$_{x}$Te ($x=0.25$). Here, $x_p = d/d_{[111]}$ is the ferroelectric displacement $d$ normalized to the unit-cell diagonal $d_{[111]}$. The pink wedge represents the region in which the Weyl nodes near the $L_1$ point are stable. As shown in the upper inset, the bulk gap at $L_1$ vanishes within the interval of lattice constants ($a$ = 6.292 $\to$ 6.308 ${\rm\AA}$) with $x_p$ fixed at 0.008. The wedge separates the topological crystalline insulator phase (TCI) from the normal insulator (NI) phase. When the ferroelectric displacement vanishes ($x_p\to 0$), the rapid shrinking of the wedge implies the merging of the two Dirac cones at all the $L$ points. This illustrates Murakami's prediction[1,2] that breaking of inversion symmetry is necessary to observe the emergence of the Weyl phase when the bulk gap is forced to close.


For the Sn content $x=$0.25, we find that Weyl nodes appear at the $L_0$ point in a very narrow range of $P$ (blue strip). Accordingly, the bulk gap at $L_0$ closes only over a much narrower interval (at the same $x_p$), as shown in the lower inset in Fig. \ref{fig:phase}A.

The $\bf k$-space orbits described by the Weyl nodes can be characterized by their radii or diameters (insets in Fig. \ref{fig:phase}B). At the point $L_0$, the projection of the orbit onto the (111) BZ surface is circular with radius $r_C = r_A$ (subscripts C and A refer to creation and annihilation points). As shown in Fig. \ref{fig:Weyl}A, the orbit at $L_0$ undulates above and below this surface with an amplitude $h$. By contrast, the orbits at $L_1$ are elliptical with major and minor diameters $d_A$ and $d_C$, respectively. In the main panel of Fig. \ref{fig:phase}B, we plot the monotonic increase of $r_C$, $d_A$, $d_C$ and $h_A$ as $x_p$ increases. Interestingly, $r_C$ and $d_C$ are closely matched throughout the range of $x_p$ explored. 


\begin{figure}[h!]
	\centering
	\includegraphics[width=9cm]{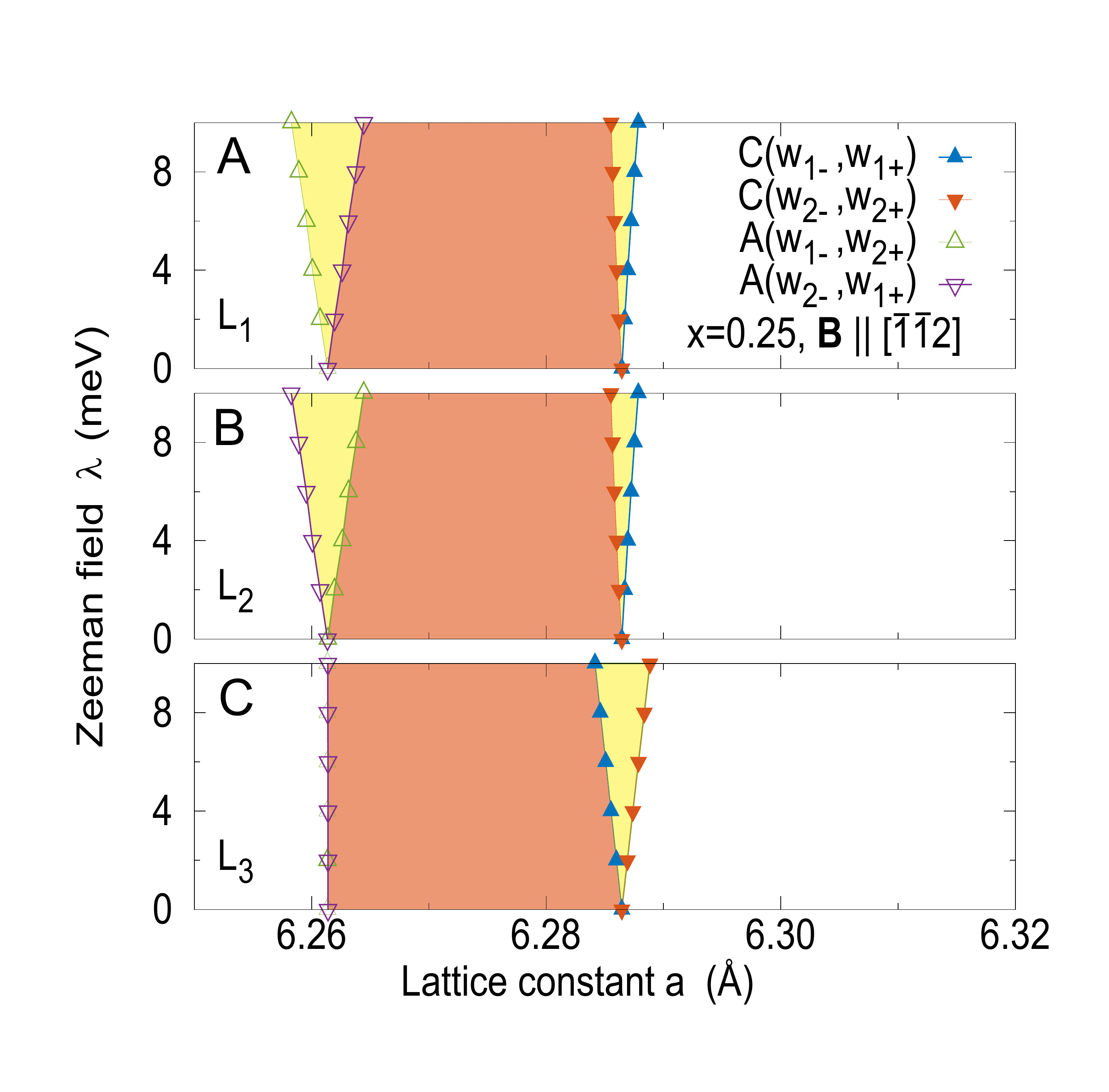}
	\caption{The phase diagram of the Weyl phase in Pb$_{1-x}$Sn$_x$Te ($x$ = 0.25) in the $a$-$\lambda$ plane with applied $\bf B\parallel$ [$\bar{1}\bar{1}2$]. Breaking of the $C_3$ symmetry about the [111] axis by $\bf B$ makes $L_3$ inequivalent to $L_1$ and $L_2$ ($L_3$ lies in the plane spanned by $\bf B$ and [111]). $x_p = 0.01$ is used. In Panel (A), a finite Zeeman field $\lambda$ splits the phase boundary for creation of Weyl nodes into two distinct boundaries ($V$-shaped wedge on the right). The annihilation boundaries (left) are similarly split. Panels (B) and (C) show the splitting for the nodes near $L_2$ and $L_3$, respectively. In each panel, the areas shaded orange (yellow) have four (two) Weyl nodes near each of the points $L_1$, $L_2$ and $L_3$.
	}
	\label{fig:Zeeman}
\end{figure}

\subsection{Effect of Magnetic Field on the Weyl Nodes}
Our calculations reveal that the Weyl node positions are highly sensitive to the time-reversal symmetry breaking effects of $\bf B$. The effects, expressed through the Zeeman energy coupling to the spins, are senstive to the direction of $\bf B$ (in finite $B$, we restore the labels of $L_1$, $L_2$ and $L_3$). 

For $\bf B\parallel$ [111], one of the two Dirac nodes appears near $L_{1}$ at a critical pressure lower than in the case with $B$ = 0 (correspondingly, its partner appears at a higher pressure). Hence the pair of nodes $w_1^{\pm}$ appears at a lower pressure than the pair $w_2^{\pm}$. The annihilation boudary is not affected if ${\bf B}$ is aligned $\parallel$[111].

By contrast, if $\bf B$ is rotated to the perpendicular direction $[\bar{1}\bar{1}2]$, the $C_3$ symmetry about the [111] axis is broken, which makes $L_3$ inequivalent to $L_1$ and $L_2$ ($L_3$ lies in the plane spanned by [111] and $\bf B$). As shown in the phase diagram in Fig. \ref{fig:Zeeman}, this leads to field-induced splittings of the phase boundaries on both sides of the Weyl phase for the nodes $L_1$ and $L_2$. In Fig. \ref{fig:Zeeman}A, the phase boundary on the right (larger $a$) representing Weyl node creation at $L_1$ splits into 2 lines as $\lambda$ increases. At the boundary on the left, a finite $\lambda$ now also affects annihilation of the Weyl nodes. The nodes $w_1^-$ and $w_2^+$ (inside the BZ volume) mutually annihilate at a pressure lower than the nodes $w_1^+$ and $w_2^-$ which lie outside the BZ. As a result, the phase boundary for annihilation splits into two lines. The effects on the phase boundaries at $L_2$ are similar, except that the Weyl nodes are swapped on the annihilation curves. The annihilation boundary for $L_3$ in Panel (C) remains unsplit (annihilation of the two Weyl pairs is simultaneous).


\begin{figure}[h]
	\centering
	\includegraphics[width=9cm]{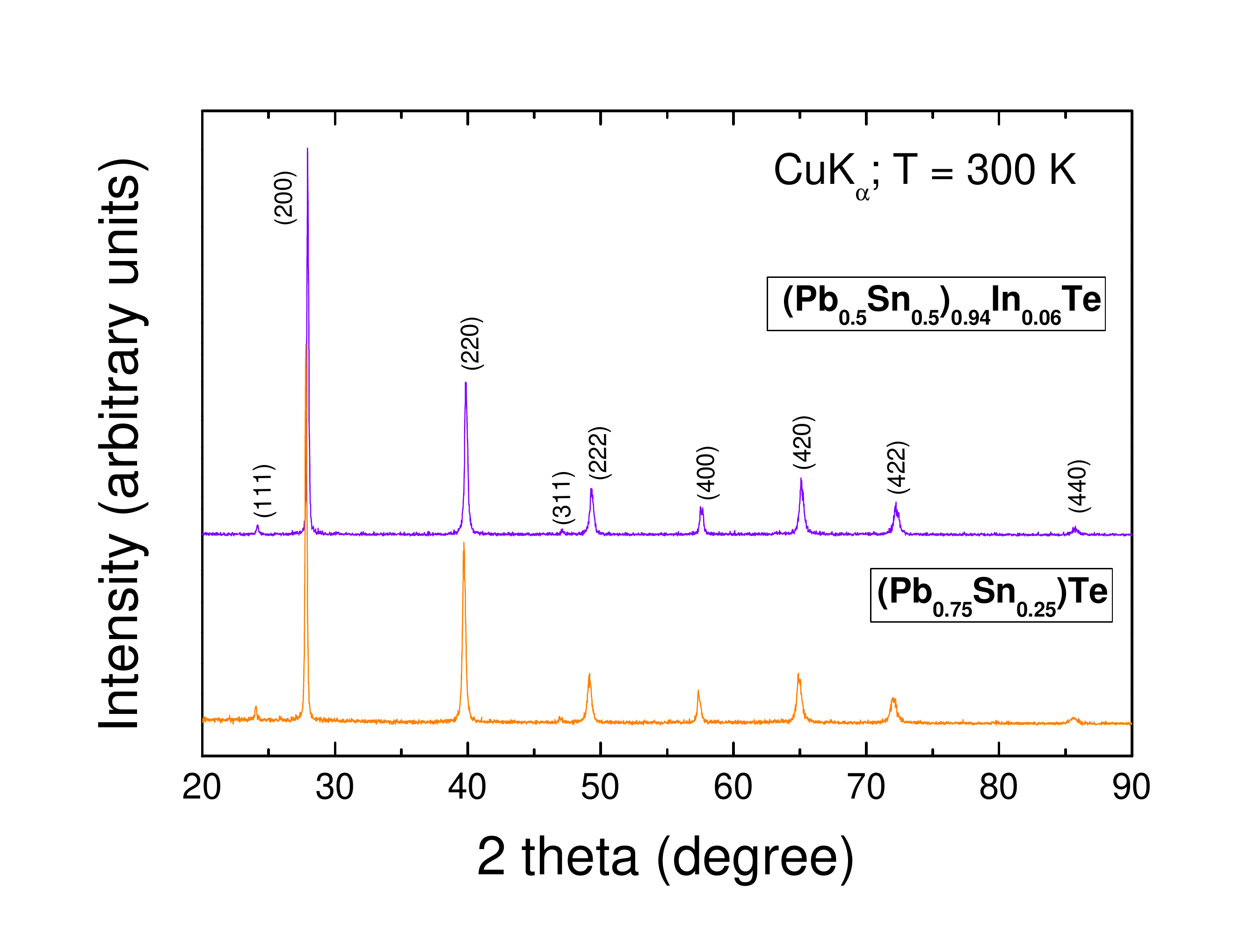}
	\caption{The X-ray diffractograms of two powdered specimens of Pb$_{1-x}$Sn$_x$Te taken from the crystal boules. The upper trace is for a boule with composition (Pb$_{1-x}$Sn$_{x}$)$_{1-y}$In$_y$Te, with $x$ = 0.5 and $y$ = 0.06). The lower trace is for a sample with composition Pb$_{1-x}$Sn$_x$Te with $x$ = 0.25.
	}
	\label{figxray}
\end{figure}

To relate to the experiment, we note that Fig. \ref{fig:Zeeman} predicts that if $P$ is fixed either just below $P_1$ or just above $P_2$ in zero $B$, the sample crosses an insulator-to-metal boundary into the Weyl phase as $B$ increases. At 5 K, this results in a very large decrease in the observed resistivity. This explains the anomalously large negative MR observed in the experiment. However, a realistic comparison with the experiment requires incorporation of the strong anisotropy of the effective $g$-factor [32,33]. Here we assumed an isotropic $g$-factor.

\begin{figure*}[t]
	\includegraphics[width=17 cm]{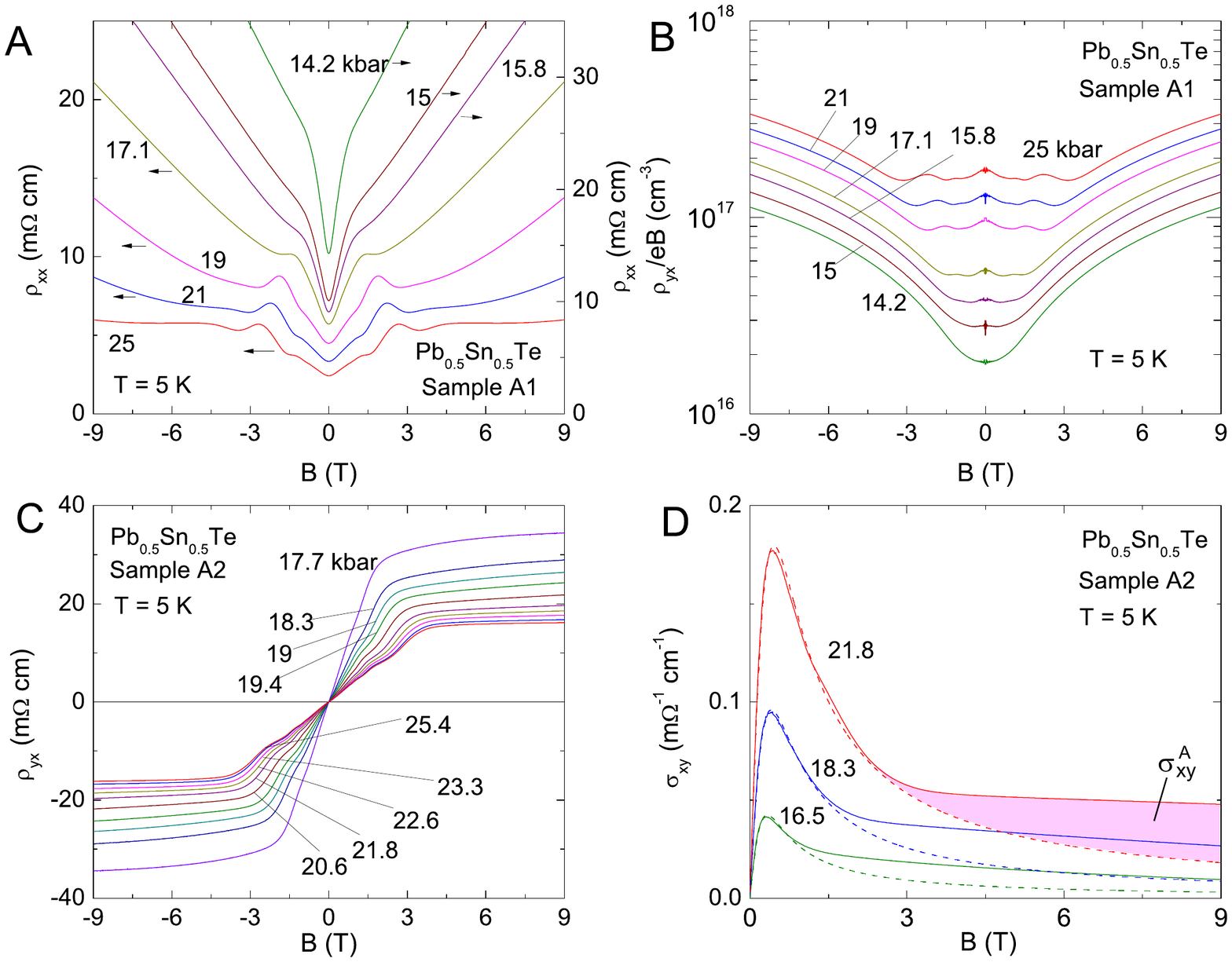}
	\caption{Supplemental data of (Pb$_{0.5}$Sn$_{0.5}$)$_{1-y}$In$_y$Te, with $y$ = 0.06 taken in Samples A1 (Panels A, B) and A2 (Panels C, D). 
		Panel (A) shows the resistivity $\rho_{xx}$ versus a transverse $B$ measured at 5 K and with $P$ fixed at values 14.2 to 25 kbar for Sample A1. At each $P$, the oscillations below $\sim 3$~T correspond to SdH oscillations (the largest peak corresponds to the $n$ = 1 LL). 
		Panel (B) plots the
		Hall resistivity divided by $Be$, $\rho_{yx}/Be$, vs. $B$ for Sample A1. At low fields (where SdH oscillations occur), the flat profile allows $\rho_{yx}/Be$ to be identified with the total hole density $n_H$. The strong increase of $\rho_{yx}/Be$ above 3 T reflects the onset of the AHE term. 
		Panel (C) plots the observed curves of $\rho_{yx}$ vs. $B$ at 5 K for Sample A2 with $P$ fixed at values above $P_1$. Instead of the conventional $B$-linear profile, $\rho_{yx}$ bends over at the ``knee'' near 2-3 T. The unusual Hall profile  suggests the appearance of an anomalous Hall conductivity that adds to the ordinary term when $B$ exceeds the knee value. 
		Panel (D) plots $\sigma_{xy}$ vs $B$ (derived from inverting $\rho_{ij}$) at 3 values of $P$ (solid curves). By fitting to Eqs.~\ref{AHEFit1}-\ref{AHEFit4}, we have separated the conventional Hall term $\sigma_{xy}^N$ (dashed curves) from the AHE term $\sigma_{xy}^A$. The latter (shown shaded in pink for the curve at 21.8 kbar) onsets as the broadened step-function $f(x)$ at $B_A$.
		\label{Sn50}
	}
\end{figure*}

\section{Field-induced anomalous Hall effect}

\subsection{Pb$_{0.5}$Sn$_{0.5}$Te}
The x-ray diffractograms recorded for two powdered specimens of two typical samples are shown in Fig. \ref{figxray}. The grown crystals are single-phased. The diffraction peaks are in very close agreement with the rocksalt crystal structure of space group \emph{Fm$\bar{3}$m}.

We focus on the topological metallic phase (regime II). As shown in Fig.~2D of the main text, the two Dirac nodes split at $P = P_1$ into two Weyl pairs at the three points $L_1$, $L_2$ and $L_3$. The size of the Fermi pockets grows under pressure, as evidenced by the evolution of SdH oscillations and the Hall density under pressure. As discussed in the main text, the system acquires an anomalous Hall contribution in finite $B$.

We first convert the measured resistivity tensor $\rho_{ij}$ into the conductivity tensor $\sigma_{ij}$ at each value of $B$. Additivity of the Hall conductivities gives
$\sigma_{xy} = \sigma_{xy}^N + \sigma_{xy}^A$. Assuming that the normal term $\sigma_{xy}^N$ is given by the conventional Drude expression, we have used the following expressions to fit the total observed Hall conductivity:
\begin{eqnarray}
	\sigma_{xy} &=& \sigma_{xy}^N + \sigma_{xy}^A,							 \label{AHEFit1}\\
	\sigma_{xy}^A &=& \sigma_{\mathrm{AHE}}^0~g(x),					 \label{AHEFit2}\\
	g(x) &=& \frac{1}{[\mathrm{e}^{-x}+1 ]},\quad ~ x = \frac{(B-B_A)}{\Delta B},				 \label{AHEFit3}\\
	\sigma_{xy}^N &=& n_H e \mu ~\frac{\mu B}{1+(\mu B)^2},						 \label{AHEFit4}
\end{eqnarray}
where $\sigma_{xy}^A$ and $\sigma_{xy}^N$ are the anomalous and Drude Hall conductivities, respectively, and $n_H$ is the carrier density. Here, the onset of the anomalous Hall effect (AHE) around $B_A$ was simulated numerically by a smooth step-function $g(x)$ with $\Delta B$ representing the width of the step, i.e., $x = (B-B_A)/\Delta B$. 

We found that Eqs. \ref{AHEFit1} to \ref{AHEFit4} provide an excellent fit to the total (observed) $\sigma_{xy}$. To highlight the anomalous contribution, we plot in Fig. \ref{Sn50}D the measured $\sigma_{xy}$ (solid curves) at the 3 pressures 16.5, 18.3 and 21.8 kbar. For comparison, we have also plotted the ordinary term $\sigma_{xy}^N$ (dashed curves) given by the Drude expression Eq. \ref{AHEFit4}. At each $P$, the difference of the 2 curves is then the anomalous term $\sigma_{xy}^A$ (simulated by $f(x)$). For the curve at 21.8 kbar, $\sigma_{xy}^A$ is the area shaded in pink.

An interesting feature inferred is that the AHE term fits the broadened step-function form $f(x)$ much better than say a $B$-linear form. The onset field $B_A$ is close to the field at which the lowest Landau level (LL) is entered. This occurs close to the knee feature in the field profiles of $\rho_{yx}$ (Fig. 3A of main text and  Fig. \ref{Sn50}C here). Further, we note that the field profile of the effective Hall number $\rho_{yx}/eB$ displays a pronounced increase above $B_A$ (Fig. 2C of the main text and Fig. \ref{Sn50}B here). The results imply that the anomalous Hall response sharply increases when $E_F$ drops into the lowest ($n=0$) LL. This could be closely related to the chiral nature of the $n=0$ LL, although there are no theoretical predictions specific to the Hall effect in the chiral LL.

From the fits, the strength of the AHE term $\sigma_{\mathrm{AHE}}^0$ can be determined. To show its behavior versus $P$, we have plotted it normalized to the hole density $p$ as $\sigma_{\mathrm{AHE}}^0/p$ vs. $P$ in Fig.~3C of the main text  (this yields the Berry curvature averaged over the FS). The corresponding results for Pb$_{0.75}$Sn$_{0.25}$Te are shown here in Fig.~\ref{Sn25}B.

A finite Berry curvature leads to an anomalous velocity ${\bf v}_A = {\bf E}\times\bf \Omega(k)$, which engenders the anomalous Hall conductivity [34]
\be
\sigma_{xy}^A = \sum_{i,\bf k} n_i({\bf k}) \Omega_{z,i}({\bf k}),
\label{eq:sxy}
\ee
where the index $i$ runs over the Weyl nodes, and $n_i$ is the occupation number in node $i$. Depending on the chirality $\chi_i$, ${\bf\Omega}_i({\bf k})$ is directed either radially inwards or outwards. Close to the node at ${\bf K}_i$, the curvature has the monopole form ${\bf\Omega}_i({\bf k})= \chi_i \Delta{\bf k}_i/|\Delta{\bf k}_i|^3$, where $\Delta{\bf k}_i = {\bf k-K}_i$ [34].

When TRS prevails (in zero $B$), the sum over Weyl nodes vanishes. In finite field, the Zeeman field $\lambda$ shifts the Fermi energy (measured from the node) in opposite directions for different signs in $\chi_i$. Because this directly affects $\Delta{\bf k}_i$, the sum in Eq. \ref{eq:sxy} yields a finite $\sigma_{xy}^A$ that grows with $\lambda$.

\subsection{Pb$_{0.75}$Sn$_{0.25}$Te}

\begin{figure*}[t]
	\includegraphics[width=17 cm]{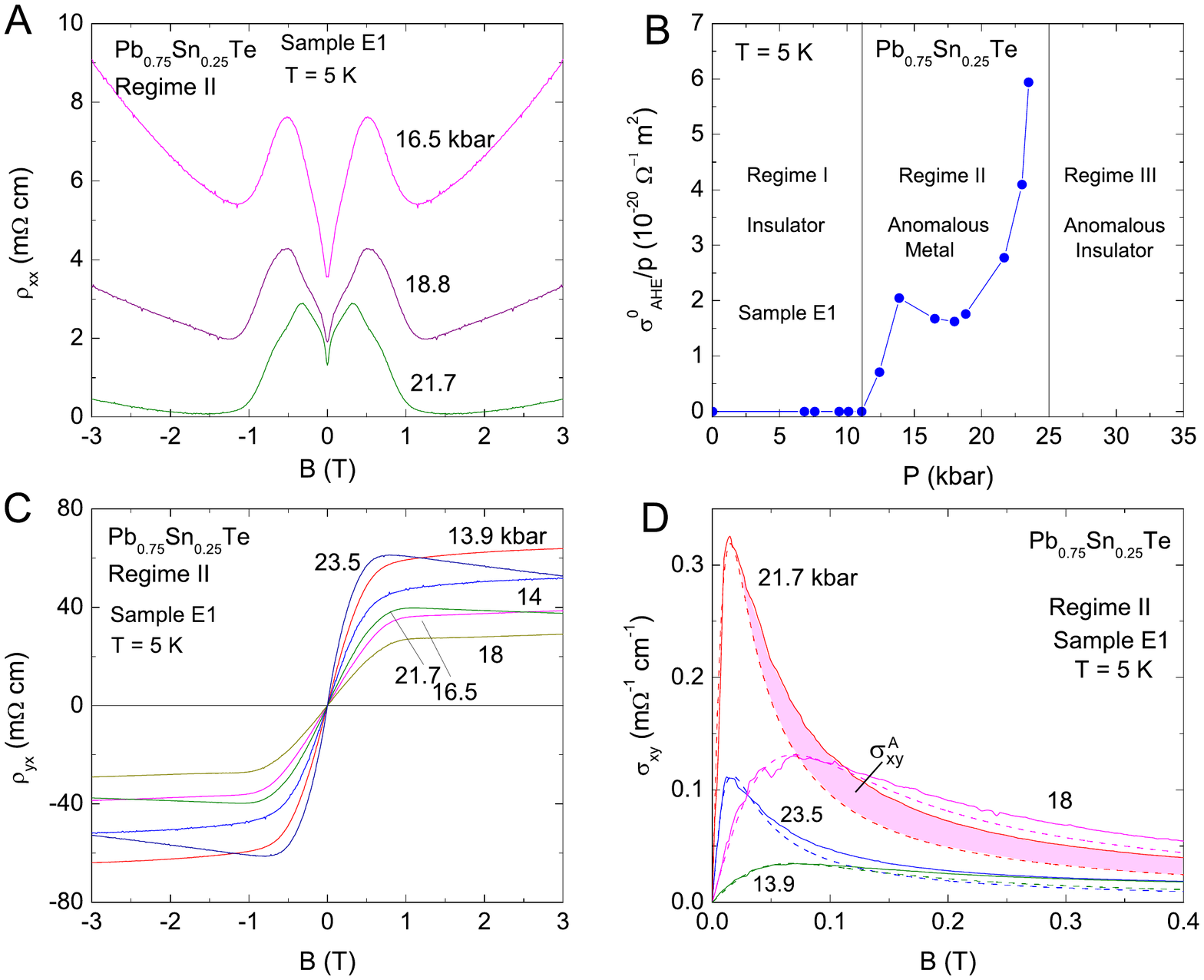}
	\caption{\label{Sn25} 
		Supplemental data of Pb$_{0.75}$Sn$_{0.25}$Te for Samples E1. 
		Panel (A) shows the resistivity $\rho_{xx}$ versus a transverse $\bf B$ measured at 5 K, with $P$ fixed at 16.5 kbar, 18.8 kbar, and 21.7 kbar (Regime II). At each $P$, the oscillations below $\sim 1$~T correspond to SdH oscillations (the largest peak corresponds to the n = 1 LL).
		Panel (B) plots the ratio of $\sigma_{xy}^A/n_H$ vs. $P$. The ratio, proportional to $<\Omega_z>$, shows a sharp increase at $P_1 \sim 11$ kbar followed by a gentler variation in the metallic phase. As $P$ approaches $P_2 \sim 25$ kbar, $\sigma_{xy}^A/n_H$ starts to diverge, signaling the appearance of the high-$P$ insulating phase.   
		Panel (C) plots the observed curves of $\rho_{yx}$ vs. $B$ at 5 K with $P$ fixed at values between $P_1$ and $P_2$ (Regime II). Instead of the conventional $B$-linear profile, $\rho_{yx}$ bends over above $\sim 1$ T, suggestive of an extraordinary contribution to $\sigma_{xy}$ at large $B$. The Hall density $n_H$ reaches its maximum value $\sim 1.58 \times 10^{16}$ cm$^{-3}$ at $P = 18$ kbar.   
		Panel (D) plots $\sigma_{xy}$ vs $B$ (derived from inverting $\rho_{ij}$) at 4 values of $P$ (solid curves). By fitting to Eqs.~\ref{AHEFit1}-\ref{AHEFit4}, we have separated the conventional Hall term $\sigma_{xy}^N$ (dashed curves) from the AHE term $\sigma_{xy}^A$. The latter (shown shaded in pink for the curve at 21.7 kbar) onsets as the broadened step-function $f(x)$ at $B_A$.
	}
\end{figure*}

Next, we briefly discuss the AHE of Pb$_{0.75}$Sn$_{0.25}$Te. In the topological metallic phase (regime II), as shown in Panels A, C of Fig.~\ref{Sn25}, the MR and Hall are essentially the same as those of Pb$_{0.5}$Sn$_{0.5}$Te in regime II except that now everything is confined in a narrower range of magnetic field. 
Therefore, the same analyses used in previous section can be employed to yield the parameters such as anomalous Hall strength $\sigma_{\mathrm{AHE}}^0$ etc. The results are $\sigma_{\mathrm{AHE}}^0 \sim 25$~$\Omega^{-1}$~cm$^{-1}$ at $p = 18$~kbar.
Unlike Pb$_{0.5}$Sn$_{0.5}$Te, in Pb$_{0.75}$Sn$_{0.25}$Te, high-$P$ insulating phase (regime III) can be achieved under experimentally accessible pressure above $\sim 25$ kbar, signaled by the divergence of $\sigma_{\mathrm{AHE}}^0/p$ shown in Panel B of Fig.~\ref{Sn25}.

The carrier density in Pb$_{0.75}$Sn$_{0.25}$Te (Sample E1) at P = 21.7 kbar is 16.7 times smaller than that of Pb$_{0.5}$Sn$_{0.5}$Te (Sample A2) at P = 25.4 kbar, and yet in both Pb$_{0.5}$Sn$_{0.5}$Te and Pb$_{0.75}$Sn$_{0.25}$Te, the maximum values of the conductivities $\sigma$ are nearly the same at $B = 0$ ($\sim 0.7$~(m$\Omega)^{-1}$~cm$^{-1}$). This is because the mobility for Pb$_{0.75}$Sn$_{0.25}$Te ($\sim 5 \pm 1 \times 10^5$ cm$^2$ V$^{-1}$ s$^{-1}$) is much larger than that of Pb$_{0.5}$Sn$_{0.5}$Te ($\sim 2.86 \times 10^4$ cm$^2$ V$^{-1}$ s$^{-1}$), allowing the system to show SdH oscillations at very low fields below $\sim 1$~T as shown in Panel A of Fig.~\ref{Sn25}, corresponding to $k_F \sim 0.005$~{\rm\AA}$^{-1}$.

\begin{thebibliography}{99}

\bibitem{Murakami1} Shuichi Murakami,
``Phase transition between the quantum spin Hall and insulator phases in 3D: emergence of a topological gapless phase,''
New Journal of Physics {\bf 9}, 356  (2007); doi:10.1088/1367-2630/9/9/356

\bibitem{MurakamiKuga} Shuichi Murakami and Shun-ichi Kuga,
``Universal phase diagrams for the quantum spin Hall systems,''
Phys. Rev. B {\bf 78}, 165313 (2008); DOI: 10.1103/PhysRevB.78.165313


\bibitem{Okugawa} Ryo Okugawa and Shuichi Murakami,
``Dispersion of Fermi arcs in Weyl semimetals and their evolutions to Dirac cones,''
Phys. Rev. B {\bf 89}, 235315 (2014); DOI: 10.1103/PhysRevB.89.235315


\bibitem{YangNagaosa} Bohm-Jung Yang and Naoto Nagaosa, 
``Classification of stable three-dimensional Dirac semimetals with nontrivial topology,''
Nat. Commun. 5:4898; doi: 10.1038/ncomms5898 (2014).

\bibitem{Wallis} Mitchell, D. L. and Wallis, R. F., 
``Theoretical energy-band parameters for the lead salts,'' 
\emph{Phys. Rev.} {\bf 151}, 581-595 (1966).

\bibitem{Dimmock} Dimmock, J.O., Melngailis, I., and Strauss, A. J., 
``Band structure and laser action in Pb$_x$Sn$_{1-x}$Te,'' 
\emph{Phys. Rev. Lett.} {\bf 26}, 1193-1196 (1966).

\bibitem{Akimov}
\bibinfo{author}{Akimov, B.~A.}, \bibinfo{author}{Dmitriev, A.~V.},
  \bibinfo{author}{Khokhlov, D.~R.} and \bibinfo{author}{Ryabova, L.~I.}
\newblock \bibinfo{title}{{Carrier Transport and Non-Equilibrium Phenomena in
  Doped PbTe and Related Materials}}.
\newblock \emph{\bibinfo{journal}{Physica Status Solidi (a)}}
  \textbf{\bibinfo{volume}{137}}, \bibinfo{pages}{9--55}
  (\bibinfo{year}{1993}). 

\bibitem{Fu} Liang Fu, 
``Topological Crystalline Insulators,'' 
\emph{Phys. Rev. Lett.} {\bf 106}, 106802 (2011).


\bibitem{Hsieh} Timothy H. Hsieh, H. Lin, J. Liu, W. Duan, A. Bansil and L. Fu, 
``Topological crystalline insulators in the SnTe material class,''
\emph{Nature Commun.} {\bf 3}, 982 (2012).


\bibitem{Story} P. Dziawa, B. J. Kowalski, K. Dybko, R. Buczko, A. Szczerbakow, M. Szot, E. Lusakowska,
T. Balasubramanian, B. M. Wojek, M. H. Bernstsen, O. Tjernberg, and T. Story, 
``Topological crystalline insulator states in Pb$_{1-x}$Sn$_x$Se,''
\emph{Nature Mater.} {\bf 11}, 1023 (2012).


\bibitem{Takahashi} Y. Tanaka, Z. Ren, T. Sato, K. Nakayama, S. Souma, T. Takahashi, K. Segawa and Y. Ando, 
``Experimental realization of a topological crystalline insulator in SnTe,''
\emph{Nature Phys.} {\bf 8}, 800 (2012).



\bibitem{Hasan} Su-Yang Xu, Chang Liu, N. Alidoust, M. Neupane, et al., 
``Observation of a topological crystalline insulator phase and topological phase transition in Pb$_{1-x}$Sn$_x$Te,'' 
\emph{Nature Commun.} {\bf 3}, 1192 (2012).

\bibitem{Madhavan}
\bibinfo{author}{Okada, Y.} \emph{et~al.}
\newblock \bibinfo{title}{{Observation of Dirac Node Formation and Mass
  Acquisition in a Topological Crystalline Insulator}}.
\newblock \emph{\bibinfo{journal}{Science}} \textbf{\bibinfo{volume}{341}},
  \bibinfo{pages}{1496--1499} (\bibinfo{year}{2013}).


\bibitem{TianLiang} Tian Liang, Quinn Gibson, Jun Xiong, Max Hirschberger, Sunanda P. Koduvayur, R.J. Cava
and N.P. Ong,
``Evidence for massive bulk Dirac fermions in Pb$_{1-x}$Sn$_x$Se from Nernst and thermopower experiments,''
Nat. Commun. 4:2696; doi: 10.1038/ncomms3696 (2013).


\bibitem{Brillson}
\bibinfo{author}{Brillson, L.~J.}, \bibinfo{author}{Burstein, E.} and
  \bibinfo{author}{Muldawer, L.}
\newblock \bibinfo{title}{{Raman observation of the ferroelectric phase
  transition in SnTe}}.
\newblock \emph{\bibinfo{journal}{Phys. Rev. B}} \textbf{\bibinfo{volume}{9}},
  \bibinfo{pages}{1547--1551} (\bibinfo{year}{1974}).


\bibitem{SI} Supplementary Materials available online

\bibitem{Nagaosa} Naoto Nagaosa, Jairo Sinova, Shigeki Onoda, A. H. MacDonald, and  N. P. Ong, 
``Anomalous Hall Effect,'' 
Rev. Mod. Phys. {\bf 82}, 1539 (2010).


\bibitem{Novak} Mario Novak, Satoshi Sasaki, Markus Kriener, Kouji Segawa, and Yoichi Ando, Phys. Rev. B {\bf 88}, 140502(R) (2013).

\bibitem{Cava} Neel Haldolaarachchige, Quinn Gibson, Weiwei Xie, Morten Bormann Nielsen, Satya Kushwaha, and R. J. Cava, Phys. Rev. B {\bf 93}, 024520 (2016).

\bibitem{Zhong} R. D. Zhong, J. A. Schneeloch, T. S. Liu, F. E. Camino, J. M. Tranquada, and G. D. Gu, 
Phys. Rev. B {\bf 90}, 020505(R) (2014).


\bibitem{Sawyer}
\bibinfo{author}{Sawyer, C.~B.} and \bibinfo{author}{Tower, C.~H.}
\newblock \bibinfo{title}{{Rochelle Salt as a Dielectric}}.
\newblock \emph{\bibinfo{journal}{Phys. Rev.}} \textbf{\bibinfo{volume}{35}},
  \bibinfo{pages}{269--273} (\bibinfo{year}{1930}).
  
\bibitem{Yamaguchi}
\bibinfo{author}{Yamaguchi, T.} and \bibinfo{author}{Takashige, M.}
\newblock \bibinfo{title}{{Key Techniques of Electric Measurements of
  Spontaneous Polarization of Ferroelectrics (in Japanese)}}.
\newblock \emph{\bibinfo{journal}{Butsuri}} \textbf{\bibinfo{volume}{66}},
  \bibinfo{pages}{603--609} (\bibinfo{year}{2011}).
  
\bibitem{Klaus}
\bibinfo{author}{M\"{o}llmann, K.-P.}, \bibinfo{author}{Herrmann, K.~H.} and
  \bibinfo{author}{Enderlein, R.}
\newblock \bibinfo{title}{{Direct observation of ferroelectric phase in
  Pb$_{1-x}$Sn$_x$Te}}.
\newblock \emph{\bibinfo{journal}{Physica B+C}} \textbf{\bibinfo{volume}{117}},
  \bibinfo{pages}{582 -- 584} (\bibinfo{year}{1983}).
 
\bibitem{KressePAW} G. Kresse and D. Joubert, From ultrasoft pseudopotentials to the projector augmented-wave method,
\textit{Phys. Rev. B} \textbf{59}, 1758 (1999).


\bibitem{Kresse96b}  G. Kresse and J. Furthm{\"u}ller, Efficiency of ab-initio total energy calculations for metals and
semiconductors using a plane-wave basis set, \textit{Comput. Mater. Sci.} \textbf{6}, 15 (1996).

\bibitem{PerdewPBEsol} J. P. Perdew, A. Ruzsinszky, G. I. Csonka, O. A. Vydrov, G. E. Scuseria, L. A. Constantin, X. Zhou, and K. Burke,Restoring the Density-Gradient Expansion for Exchange in Solids and Surfaces,
\textit{Phys. Rev. Lett.} \textbf{100}, 136406 (2008).



\bibitem{Kim2010} Yoon-Suk Kim,  Martijn Marsman, Georg Kresse, Fabien Tran, and Peter Blaha,
Towards efficient band structure and effective mass calculations for III-V direct band-gap semiconductors,
\textit{Phys. Rev. B} \textbf{82}, 205212 (2010).


\bibitem{Mariano1967} A. N. Mariano and K. L. Chopra, \textit{Appl. Phys. Lett.} \textbf{10}, 282 (1967).

\bibitem{Mostofi} A. A. Mostofi, J. R. Yates, G. Pizzi, Y.-S. Lee, I. Souza, D. Vanderbilt, and N. Marzari, An updated version of wannier90: A tool for obtaining maximally-localised Wannier functions, \textit{Comput. Phys. Commun.} \textbf{185}, 2309 (2014).





\bibitem{Tsu1968} R. Tsu, W. E. Howard, and L. Esaki, \textit{Phys. Rev.} \textbf{172}, 779 (1968).

\bibitem{VanderbiltPRB2014} J. Liu and D. Vanderbilt., Weyl semimetals from noncentrosym-metric topological insulators, \textit{Phys. Rev. B} \textbf{90}, 155316 (2014).


\bibitem{Dimmock} J. O. Dimmock, \textit{Physics of Semimetals and Narrow-Gap Semiconductors}, edited by D. L. Carter and R.T. Bate (Pergamon, New York, 1971).

\bibitem{RidolfiPRB2015} E. Ridolfi, E. A. de Andrada e Silva, and G. C. La Rocca, Effective $g$-factor tensor for carriers in IV-VI semiconductor quantum wells, \textit{Phys. Rev. B} \textbf{91}, 085313 (2015).


\bibitem{Nagaosa2003}
Zhong Fang \etal, 
``The anomalous Hall effect and magnetic monopoles in momentum space,''
{\emph Science} {\bf 302}, 92 (2003)	

\end{thebibliography}
\end{document}